\documentclass[12pt]{article}
\newlength{\abstractwidth}
\pdfoutput=1

\usepackage{amsmath}
\usepackage{epsf}
\usepackage{color}
\usepackage{graphicx}
\usepackage{amsfonts, amssymb}

\newcommand{\beq}{\begin{equation}}
\newcommand{\eeq}{\end{equation}}
\newcommand{\bsp}{\begin{split}}
\newcommand{\esp}{\end{split}}
\newcommand{\T}{\mathcal{T}}
\newcommand{\M}{\mathcal{M}}

\newcommand{\N}{\mathcal{N}}

\numberwithin{equation}{section}

\newcommand{\xb}{\textbf{x}}
\newcommand{\rar}{\rightarrow}
\newcommand{\cD}{{\cal D}}
\newcommand{\cL}{{\cal L}}
\newcommand{\cO}{{\cal O}}

\begin{document}
\setcounter{footnote}{0}
\begin{flushright}
\end{flushright}

\bigskip

\begin{center}

{\LARGE \bf Universality and exactness of Schr\"odinger} \\
 \medskip
{\LARGE \bf geometries in string and M-theory}

\vskip .4in

{\large \bf   Per Kraus and Eric Perlmutter}

\vskip .2in

{ \sl Department of Physics and Astronomy }\\
{\sl University of California, Los Angeles, CA 90095, USA}\\
{\tt pkraus@ucla.edu, perl@physics.ucla.edu }

\end{center}

\vskip .2in

\begin{abstract}

\vskip 0.1in

We propose an organizing principle for classifying and constructing Schr\"odinger-invariant solutions within string theory and M-theory, based on the idea that such solutions represent nonlinear completions
of linearized vector and graviton  Kaluza-Klein excitations of AdS compactifications. A crucial simplification, derived from the symmetry of AdS, is that the nonlinearities appear only quadratically.  Accordingly, every AdS vacuum admits infinite families of Schr\"odinger deformations parameterized by the dynamical exponent $z$.  We exhibit the ease of finding these solutions by presenting three new constructions: two from M5 branes, both wrapped and extended, and one from the D1-D5 (and S-dual F1-NS5) system. From the boundary perspective, perturbing a CFT by a null vector operator can lead to nonzero $\beta$-functions for spin-2 operators; however, symmetry restricts them to be at most quadratic in couplings.

This point of view also allows us to easily prove nonrenormalization theorems: for any Sch($z$) solution of two-derivative supergravity constructed in the above manner, $z$ is uncorrected to all orders in higher derivative corrections if the deforming KK mode lies in a short multiplet of an AdS supergroup. Furthermore, we find infinite classes of 1/4 BPS solutions with 4-,5- and 7-dimensional Schr\"odinger symmetry that are exact.

\end{abstract}

\newpage

\contentsline {section}{\numberline {1}Introduction}{2}
\contentsline {section}{\numberline {2}Infinite families of Schr\"odinger solutions from every AdS}{4}
\contentsline {subsection}{\numberline {2.1}Review and motivation}{5}
\contentsline {subsection}{\numberline {2.2}In the bulk: AdS perturbation theory}{9}
\contentsline {subsection}{\numberline {2.3}On the boundary: conformal perturbation theory}{12}
\contentsline {subsection}{\numberline {2.4}A corollary: consistent truncations with massive KK modes}{15}
\contentsline {section}{\numberline {3}Quantum/string corrections}{16}
\contentsline {subsection}{\numberline {3.1}Uncorrected Schr\"odinger solutions from short multiplets}{17}
\contentsline {subsection}{\numberline {3.2}At $O(\alpha'^3)$, explicit nonrenormalization of Sch$_5$}{19}
\contentsline {section}{\numberline {4}New Schr\"odinger solutions}{23}
\contentsline {subsection}{\numberline {4.1}Sch$_7$ from M5 branes}{23}
\contentsline {subsection}{\numberline {4.2}Sch$_3$ from D1-D5 and F1-NS5 branes}{25}
\contentsline {subsection}{\numberline {4.3}Sch$_4$ from wrapped M5 branes}{28}
\contentsline {section}{\numberline {5}Discussion} {30}
\contentsline {section}{\numberline {A}More non-relativistic solutions from M5 branes}{31}
\contentsline {section}{\numberline {B}Details of Sch$_3$ solutions from D1-D5}{34}
\contentsline {section}{\numberline {C}Details of Sch$_4$ solutions from wrapped M5} {40}

\setcounter{equation}{0}

\section{Introduction}

Since the early days of bottom-up Galilean holography \cite{Balasubramanian:2008dm, Son:2008ye}, various techniques have been developed to embed Schr\"odinger solutions into string theory and M-theory. The TsT transformation, or Null Melvin Twist, has been used to transform the asymptotics of various black brane solutions \cite{Herzog:2008wg, Adams:2008wt, oz, Imeroni:2009cs, Adams:2009dm, Hartong:2010ec, Banerjee:2011jb}, and zero temperature solutions have been embedded into consistent truncations of type IIB and eleven-dimensional supergravities that retain massive vector modes, thereby making contact with the original bottom-up approach \cite{mald, Gauntlett:2009zw, Cassani:2010uw, Liu:2010sa, Gauntlett:2010vu, Bena:2010pr, Cassani:2010na, O'Colgain:2009yd}. More general solutions, formed directly in $D=10,11$, engineer non-relativistic conformal symmetry geometrically, drawing a connection between harmonic modes on internal manifolds and zero temperature Schr\"odinger solutions with infinite spectra of $z$, e.g. \cite{hartnoll, gaunt2, gaunt, gaunt3, bobev, Ooguri:2009cv, jeong, Singh:2010rt}.

For extremal solutions, the construction of \cite{gaunt} is the most general to date. Building on their own work \cite{gaunt2} and followed by \cite{gaunt3}, the authors' framework subsumes all previously found zero temperature embeddings based on null deformations of the near-horizon D3 and M2-brane solutions, realizing them as members of infinite families of Schr\"odinger solutions. The spectrum of $z$ is determined by the spectrum of harmonics on Sasaki-Einstein manifolds. The geometric nature of these solutions suggests an extension to other deformations of AdS vacua with some generality.

Given these developments, the outstanding question is no longer how top-down Schr\"odinger solutions can be found, but instead, what the fundamental principle is that allows us to explain their existence and classify them. In this work, largely inspired by \cite{hartnoll, gaunt, bobev}, we attempt to answer the following question: given an AdS$ \, \times\, \M$ solution, can it always be deformed to have Schr\"odinger symmetry? The answer, we argue, is yes.

We begin by elucidating and extending the constructions of \cite{hartnoll, gaunt, bobev}, which stand in direct correspondence to linearized Kaluza-Klein vectors and gravitons of AdS compactification spectra. In fact, we show that this is more than a parallel. To every such spin-2 excitation, there exists a nonlinear solution with Schr\"odinger symmetry that is obtained by ``turning on'' the KK mode; and in rather general circumstances, there are Schr\"odinger solutions that involve both vector and spin-2 excitations which couple only at quadratic order.

The obvious puzzle is why linearized solutions can be extended to nonlinear order in such a simple way. We use perturbation theory around a generic AdS$ \, \times\, \M$ vacuum to show that the nonlinearities are constrained by symmetry to arise only at second order. Scale invariance consistent with the first-order solutions determines the geometry to be of Schr\"odinger form. Putting this all together, for every AdS$ \, \times\, \M$ vacuum in string and M-theory, there are Schr\"odinger solutions in direct correspondence with vector and graviton KK modes of the compactification spectra on $\mathcal{M}$. The most general Schr\"odinger solutions, which exist when the harmonic spectra on $\M$ admit solution of a simple set of Laplace equations, superpose these modes. 

In this sense, the existence of Schr\"odinger solutions is universal, and their construction, mechanical. We support our arguments by providing three new categories of solutions explicitly, with three-, four- and seven-dimensional Schr\"odinger symmetry. This includes both supersymmetric and non-supersymmetric solutions.

We also provide an argument on the CFT side. The dual version of the quadratic truncation of the AdS perturbation theory is that the $\beta$-function equations for all deforming operators truncate at second order in the couplings, modulo some issues regarding the effect of multi-trace operators.  When perturbing the theory by a source for a null
vector operator, scale invariance is preserved provided that no marginal
spin-2 operator appears in the OPE of two vector operators.  Further parallels can be drawn by considering the Wilsonian $\beta$-functions equations for a
tower of spin-2 operators.  Their $\beta$-function equations are again
constrained by symmetry to be quadratic in the couplings, and
the non-relativistic fixed point is easily solved in terms of operator dimensions and constants arising from OPEs between vector operators. This is particularly useful in light of recent work in setting up the precise holographic dictionary for Schr\"odinger gauge/gravity duality, which focused primarily on solutions to massive vector theories \cite{Costa:2010cn, Guica:2010sw}. But the most general embeddings of the type analyzed in this work do not fit into that framework:  there are not only massive vectors but massive gravitons as well, dual to symmetric tensor operators with nontrivial Wilsonian $\beta$-functions.

This perspective has powerful implications for the existence of the Schr\"odinger backgrounds beyond the two-derivative approximation, first considered in \cite{adams}. In particular, we recall that conformal dimensions of operators in short multiplets of an AdS supergroup, which are determined via harmonics of $\M$, are unrenormalized by quantum and stringy corrections. This implies that any parameter of a Schr\"odinger solution that derives from the dimensionless AdS mass of a KK mode can be so protected. This is the case for the dynamical exponent, $z$. Furthermore, when we perturb the exact AdS vacua that arise in the near-horizon limit of M2, D3 and M5 branes, certain infinite families of the resulting Schr\"odinger solutions are uncorrected to all orders in a higher-derivative expansion of the action. These vacua preserve only eight Poincar\'e supersymmetries, and constitute a new contribution to the small set of exact solutions of string and M-theory. We back this claim with an explicit calculation showing that one such class of Sch$_5$ solutions of type IIB is unrenormalized to $O(\alpha'^3)$, inclusive of \textit{all} correction terms, known \cite{paulos} and unknown.

We briefly note that all of our arguments -- in particular, the truncation of the perturbation theory at quadratic order -- neglect possible effects of multi-trace operators, which appear to play no role in the present context in the large $N$ limit. The existence of classical Schr\"odinger solutions is apparently robust against effects of operators of spin greater than two. Scale-invariance may be lost upon including $1/N$ effects, but we do not consider this issue here.

The rest of this paper is organized as follows. Section 2 reviews the Schr\"odinger solutions of type IIB obtained by a D3-brane deformation, and generalizes them with the perturbation theory arguments detailed above, both in the bulk and on the boundary. We also consider the implications for embedding Schr\"odinger solutions in consistent truncations with massive modes. Section 3 posits conditions for nonrenormalization. Section 4 presents the new Schr\"odinger constructions, details of which are in the appendices, and section 5 concludes with a discussion.

\textbf{Note added, v2:} Appendix A now presents another infinite family of solutions with seven-dimensional Schr\"odinger symmetry obtained through deformation of the extended M5 brane geometry that does \textit{not} lie in obvious correspondence with massive KK modes found in the compactification on AdS$_7 \times M_4$, as well as a single solution with six-dimensional Lifshitz symmetry with the same general structure as those Lifshitz solutions of \cite{koush, gauntlif}.

\section{Infinite families of Schr\"odinger solutions from every AdS}

In reviewing the type IIB solutions of \cite{hartnoll, gaunt, bobev}, we emphasize how their structure parallels that seen in the harmonic analysis of the Kaluza Klein spectrum on AdS$_5 \times SE_5$. We then show that infinite towers of Schr\"odinger solutions are guaranteed to exist, for any AdS vacuum, and are governed by the structure of the KK towers of massive vectors and gravitons. For convenience, we use the notation and supergravity conventions of \cite{gaunt}, the latter of which are found in \cite{IIB}.

\subsection{Review and motivation}

The type IIB solutions of \cite{gaunt} are\footnote{Note a minor sign error in \cite{gaunt2, gaunt} of the $\star_{CY_3}dC$ term, where we, and they, use the convention $\epsilon_{+-x_1x_2y_1\ldots y_6}=+\sqrt{-g}$, where the $y_i$ are coordinates on $CY_3$.}
\beq
\begin{split}
ds^2 &= \Phi^{-1/2}\left(2dx^+ dx^- + h(dx^+)^2 + 2Cdx^+ + d\textbf{x}^2\right) + \Phi^{1/2}ds^2(CY_3)\\
F_5 &= dx^+ \wedge dx^- \wedge dx_1 \wedge dx_2 \wedge d\Phi^{-1} + \star_{CY_3}d\Phi\\
&\quad -dx^+\wedge \left(d(\Phi^{-1}C) \wedge dx_1 \wedge dx_2-\star_{CY_3}dC\right)\\
G_3 &= dx^+ \wedge W\\
\end{split}
\eeq
This solution is a deformed version of D3-branes sitting at the  tip of a $CY_3$ cone.
$\Phi$ and $h$ are functions; $C$ is a one-form; and $W$ is a  complex two-form; all of which are defined on $CY_3$. The axion-dilaton is set to zero, and $\textbf{x}=(x_1,x_2)$. The  field equations evaluated on this ansatz reduce to
\beq\label{eqns1}
\begin{split}
\nabla^2_{CY_3}\Phi=0\\
d\star_{CY_3}dC=0\\
dW=d\star_{CY_3}W=0\\
\nabla^2_{CY_3}h = -|W|^2_{CY_3}
\end{split}
\eeq
where $|W|^2_{CY_3}$ is the real square of $W$ with indices contracted with the  $CY_3$ metric. The first of these equations is the usual harmonic condition on $\Phi$, which we use to zoom into the near-horizon region by making the substitution $\Phi = r^{-4}$, and writing the $CY_3$ as a cone over a Sasaki-Einstein space $SE_5$ as
\beq
ds^2(CY_3) = dr^2 + r^2ds^2(SE_5).
\eeq

Upon substitution, we can satisfy the remaining three equations by taking
\beq\label{fields}
\begin{split}
C &= r^{z-2}\beta\\
W &= d(r^z\sigma)\\
h &= r^{2z-2}q\\
\end{split}
\eeq
where $q$, $\beta$ and $\sigma$ are a function, real one-form, and complex one-form, respectively, on $SE_5$. The powers of $r$ follow from insisting upon anisotropic scale invariance under the transformation
\beq\label{scaling}
r \rar r'=\lambda r \, , \quad x^+ \rar {x^{+}}'= {x^+ \over \lambda^z}\, ,\quad x^- \rar {x^{-}}'={x^- \over \lambda^{2-z}}\, ,\quad \xb \rar \xb'= {\xb \over \lambda}
\eeq
and the solution now reads
\beq\label{soln}
\begin{split}
ds^2 &=  r^{2z}q(dx^+)^2 + 2r^{z}dx^+\beta+ r^2\left(2dx^+ dx^-  + d\textbf{x}^2\right) + \frac{dr^2}{r^2}+ds^2(SE_5)\\
F_5 &= 4r^3dx^+ \wedge dx^- \wedge dx_1 \wedge dx_2 \wedge dr + 4 Vol_5\\
&\quad - d\left(r^{z+2}dx^+\wedge dx_1 \wedge dx_2\wedge \beta\right)+dx^+\wedge \left(r^{z-1}dr\wedge \star_{5}\beta + (z-2)r^z\star_5d\beta\right) \\
G_3 &= d\left(r^zdx^+ \wedge\sigma\right) \\
\end{split}
\eeq
where $Vol_5$ and $\star_5$ are the invariant volume form and Hodge star operator on $SE_5$, respectively. The Schr\"odinger symmetry is now manifest, and the  field equations reduce to
\beq\label{fieldeqns}
\begin{split}
&\Delta_5\beta = z(z-2)\beta\, , \quad \text{where} \quad d\star_5\beta=0\\
&\Delta_5\sigma = z(z+2)\sigma\, , \quad \text{where} \quad d\star_5\sigma=0\\
&\nabla^2_5q + (2z-2)(2z+2)q = -z^2|\sigma|^2_5 - |d\sigma|^2_5
\end{split}
\eeq
with $-\nabla^2_5$ and $\Delta_5$ as the Laplace operator on functions and one-forms on $SE_5$, respectively.

The most general solution in which all of these fields are turned on is subject to the consistency of all equations. First, one should think of $z$ as determined by the vector harmonic. Then the solution to the third equation, an inhomogeneous Laplace equation, will only exist if the quadratic vector source, when expanded in a basis of scalar harmonics on $\M$, does not source the homogeneous mode. It was proven in \cite{gaunt} that for $z=2$, these equations can be solved for any $SE_5$, and that for $SE_5=S^5$, these equations can in fact be solved for all $z$. \\

Having reviewed these  solutions, we now make some observations. The structure of this ansatz, and the form of  the reduced field equations, is almost  identical to that which arises in solving for linearized fluctuations on AdS$_5\times SE_5$.  The only difference, which we'll return to shortly, is the quadratic
nonlinearity appearing in the final equation of (\ref{eqns1}).
Take the $\sigma \neq 0$ solutions, for example. We have written the three-form flux as
\beq
G_3 = dA_2 = d\left(r^zdx^+ \wedge\sigma\right)
\eeq
in order to emphasize the structure of the complex two-form gauge field $A_2$: it has one leg along $SE_5$, and one leg along the five noncompact directions that survive the compactification. The latter becomes the gauge field in $d=5$, to wit, the Schr\"odinger gauge field of an effective $d=5$ massive vector model,
\beq
A_1^{Sch} = r^zdx^+
\eeq
Furthermore, we found that $\sigma$ must be a transverse vector harmonic on $SE_5$, and $z(z+2)$ is identified with its eigenvalue. The spectrum of co-closed one-forms on $SE_5$ has eigenvalues $\Delta_5 \geq 8$, where this bound is saturated when $\sigma$ is dual to a Killing vector on $SE_5$. This gives a lower bound\footnote{Here and henceforth, we ignore negative $z$ consistent with the field equations.}
\beq
z \geq 2
\eeq

But  this is exactly how the tower of KK vectors with a diagonal field equation\footnote{Henceforth called ``diagonal'' vectors.} arises in the harmonic analysis of linearized fluctuations: one expands the internal part of the complex two-form gauge field in harmonics on $SE_5$, plugs into the linearized field equations subject to a particular gauge choice, and the equation for the fluctuations $A_1$ becomes the eigenvalue equation of vector harmonics. The de Donder-Lorenz gauge choice forces these vector harmonics to be transverse, which is to say, co-closed on $SE_5$: this is just the condition $d\star_5\sigma=0$ that we have imposed above. The resulting definition of $z$ is simply that of a five-dimensional massive vector model,
\beq\label{kk}
(mL)^2 = z(z+2)
\eeq
where $(mL)^2$ is here the dimensionless AdS mass of the KK vector. Notice that the ansatz is linear in $W$, as it is in all of the deformation fields, and the axio-dilaton need not be turned on because it does not figure into the Kaluza-Klein vector spectrum.

There is another field equation to be obeyed, namely the Einstein equation given in (\ref{fieldeqns}). Notice that when $\sigma=0$, the equation is simply that of a massive spin-2 excitation on AdS$_5 \times SE_5$, where $q$ is a scalar harmonic with eigenvalue $4(z^2-1)$. Its harmonic level determines the mass of the spin-2 field. One should think of expanding $h$ in $SE_5$ scalar harmonics as
\beq
h = \sum_kh_{\mu\nu}^kY^k(SE_5)
\eeq
thus identifying the spin-2 field as
\beq
h_{++}=r^{2z}
\eeq
Notice that $h_{++}$ is a transverse traceless mode because of the null Killing vector of the Schr\"odinger metric; this, too, parallels the gauge choice of the compactification.

Evidently, taking $\sigma \neq 0$ induces a quadratic correction to this equation; the solution is given as a linear superposition of scalar harmonics on $SE_5$, as determined by the $SE_5$ dependence of the source term.\footnote{There is an exception to this when $z=2$, for which the $\sigma$-dependence of the spin-2 equation reduces to a constant. In this case, the most general solution is $g_{++} = qr^{4}$, where $q = a + bY(SE_5)_{12}$ for some constants $a,b$ and $Y(SE_5)_{12}$ the scalar harmonic with eigenvalue 12 (if it exists). The choice $b=0$ does not turn on a spin-2 field because there is no constant scalar harmonic on $SE_5$ other than zero. This is, in fact, the near-horizon limit of the TsT-transformed extremal D3-brane.} Thus, the $\sigma \neq 0$ solutions require not only a vector perturbation to AdS$_5 \times SE_5$, but a tower of spin-2 perturbations whose amplitudes are proportional to the square of the vector harmonic.

A similar explanation holds for the $\beta \neq 0$ solutions: these make use of the vectors that descend from a mixture of the ten-dimensional metric and four-form RR gauge field\footnote{Henceforth called ``mixed'' vectors.}. The five-form ansatz in (\ref{soln}) has various $\beta$ terms which combine to make a self-dual $F_5$. Focusing on the term of the form
\beq
F_5=dA_4 \, \supset \, d\left(r^{z+2}dx^+\wedge dx_1 \wedge dx_2\wedge \beta\right)
\eeq
we see that we have turned on a component of the four-form gauge field with \textit{three} legs along the extended directions. Nevertheless, this is precisely the form of the ansatz needed to turn on the mixed KK vectors: in the linearized KK analysis, these components are algebraically eliminated in favor of a component with only one extended index. (See discussion above equation (2.17) of \cite{van nieu} when $SE_5 = S^5$.) These solutions will have $z \geq 4$, and the massive vector relation (\ref{kk}) still holds.\footnote{In this and other constructions using mixed vectors, there are two such towers. Because the eigenvalue equation that determines $z$ is quadratic, it is satisfied by two choices of $z$. The greater choice turns on a vector in the ``upper'' branch; the lesser choice, the ``lower'' branch. Typically, the values of $z$ corresponding to the lower branch are negative, and we will ignore them for physical reasons.}

One should keep in mind that the spectrum of $z$ is bounded from above by the validity of the supergravity approximation, just as the masses of KK modes are bounded from above by the string scale: in particular, $z \lesssim \lambda^{1/4}$ with $\lambda$ large.\\


Upon taking $SE_5 = S^5$ \cite{van nieu}, the above exposition becomes rather transparent: the eigenvalue spectra of the Laplace operators are integer and stand in the right relation to give a spectrum of Schr\"odinger solutions with integer $z$. A graph of the vector spectrum about AdS$_5 \times S^5$ then supplies a visual representation of the associated Schr\"odinger deformations, provided in Figure 1. \\
\begin{figure}[h!]
\begin{centering}
  \includegraphics[scale=.47]{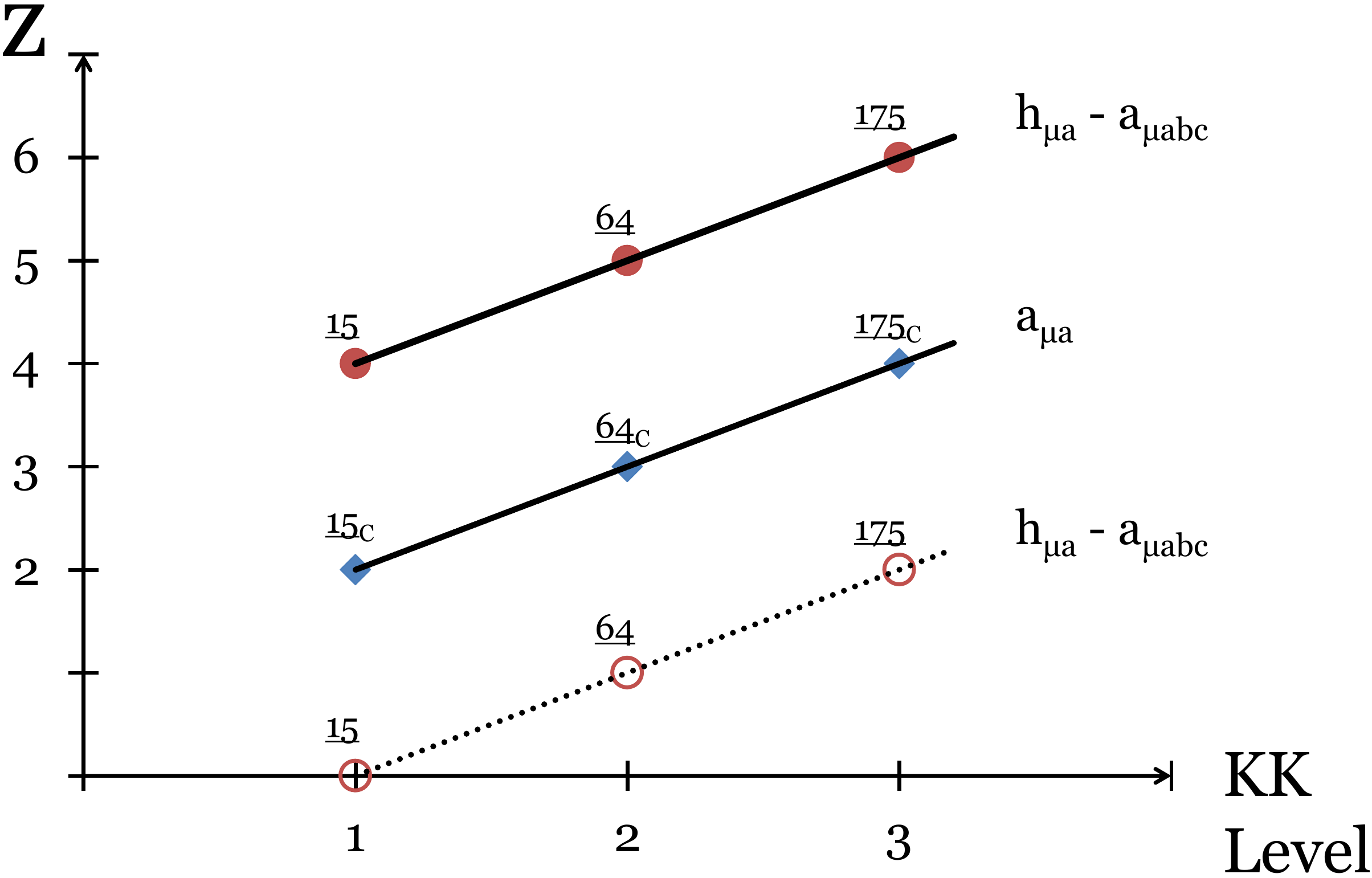}\\
\end{centering}
\caption{\small The lower states of the spectrum of Sch$_5(z)$ solutions formed from Kaluza-Klein vector deformations of AdS$_5 \times S^5$. This is an adaptation of Figure 1 in \cite{van nieu} which depicts the vector mass spectrum, here converted via the massive vector relation $(mL)^2_{KK}=z(z+2)$. The $SO(6)$ representation of each vector, listed above each point, partly determines the degeneracy of solutions. At right are the $D=10$ origin of fields, where $\mu$ lies along the noncompact directions and $(a,b,c)$ lie on $S^5$. The middle branch of solutions, formed from diagonal vectors descending from the complex two-form, also requires nonzero Kaluza-Klein gravitons not pictured. The lower branch of mixed vectors, while present in the spectrum, gives solutions with negative values of $z$; hence, it appears dotted and unfilled. For details, see the main text.}
\end{figure}\\
\textbf{Generalization to all AdS vacua}\\

The structure of the field equations \eqref{fieldeqns}, and our analysis thereof, applies equally to the M2-brane solutions constructed in \cite{gaunt}. One might ask whether there is anything special about these two sets of solutions, with an eye toward generalizing this structure to solutions built around any AdS$\times \mathcal{M}$ vacuum.

First, as we will soon note, the fact that the vectors couple quadratically to the gravitons is a completely general phenomenon that derives solely from the Lorentz symmetry of AdS. Combined with the Kaluza-Klein perspective above, it is clear that one can generalize this construction to null deformations of all AdS spacetimes. On the other hand, notice that the mixed vectors do not couple to the gravitons: that is, one might have expected a quadratic source term in $\beta$ to appear in the third (Einstein) equation in \eqref{fieldeqns}. This absence is likely due to the supersymmetry of the associated AdS solutions, and we remark on this further in section 3.

 Summarizing, to each KK vector of an AdS$_{d+1} \times \mathcal{M}$ compactification, one can associate a set of reduced field equations analogous to \eqref{fieldeqns} with at most quadratic source terms. The dynamical exponent $z$ is given precisely by the relation for an effective $d$-dimensional massive vector model,
\beq\label{kkd}
(mL)^2_{KK} = z(z+d-2)
\eeq
where the mass is simply the AdS mass of the KK vector: by construction, the same differential operator of $\mathcal{M}$ that gives the spectrum of vector masses is that which gives $z$. If these equations can be solved, one has a solution with Sch$_{d+1}(z)$ symmetry where the Schr\"odinger vector is \textit{identified} with the linearized KK fluctuation. The metric deformation is induced by a tower of massive spin-2 fields that together solve an inhomogeneous Laplace equation with a quadratic source in the vector harmonic.

One can also utilize the spin-2 fields alone to obtain a Schr\"odinger-symmetric solution, where $z$ is determined by the effective mass of the spin-2 field as
\beq\label{kkgd}
(mL)^2_{KK} = \left[d+2(z-1)\right]2(z-1)
\eeq
(We will justify this relation for all $d$ later in this section.) Finally, in all of these constructions, the degeneracy of Schr\"odinger solutions is given by the degeneracy of $\mathcal{M}$ harmonics utilized in the solution.

Analysis of the KK spectrum is based on solving linearized field equations, while the Schr\"odinger solutions
represent solutions of the fully nonlinear theory.
In these terms, an obvious question is: how does a solution at linearized level become elevated to a full, nonlinear solution?  We now explain how this occurs, first from the point of view of the
bulk field equations, and subsequently from the point of view of the holographically dual boundary
theory.

\subsection{In the bulk: AdS perturbation theory}

The message of the above remarks is that one should think of the nonlinear solution as a quadratic extension of a linearized solution around AdS$_5 \times SE_5$. Thus, our task is to explain why this perturbation theory truncates at second order, and to generalize this fact.

Still referring to the D3-brane case, our strategy will be to initiate a perturbation theory around the AdS solution by seeding it with a linearized vector fluctuation, where the three-form flux has a leg in the $x^+$ direction. The fluctuation has some definite weight under the symmetry transformations of the background. Going to higher order in the perturbation theory, the symmetries of this first order solution must be preserved at each order. Furthermore, the possible terms that can get generated must be covariant with respect to the symmetries of the background. In general, this perturbation series will not truncate at finite order; in particular, if one can combine positive powers of the first-order perturbations to form a singlet under all background symmetries, then one will expect this singlet to appear in equations at arbitrary order.  Simplification occurs if, for any one symmetry of the background, there are no singlets.

In the case at hand, the Lorentz symmetry of AdS, acting in these coordinates as
\beq\label{lorentz}
x^+ \rar {x^{+}}'={x^+ \over \kappa}\, ,\quad x^- \rar {x^-}'=\kappa  x^-
\eeq
strongly constrains the possible terms that can be generated beyond linear order. Because we seed the perturbation theory with a field with a leg along the null $x^+$ direction, and with \textit{no} legs along the $x^-$ direction, this field has weight one under the Lorentz boost. That means that at $n^{th}$ order, the only possible terms that can be generated are of weight $n$. But because the lone symmetric tensor field in our theory is the metric, which has two indices, this perturbation theory will necessarily truncate at quadratic order: there are no possible terms one can write down which turn on a field with Lorentz weight three.

Demonstrating explicitly with the $W \neq 0$ solution based on diagonal vectors, we seed the perturbation via the complex three-form flux, as
\beq
\begin{split}
ds^2 &= r^2\left(2dx^+ dx^-  + d\textbf{x}^2\right) + \frac{dr^2}{r^2}+ds^2(SE_5)\\
F_5 &= 4r^3dx^+ \wedge dx^- \wedge d\textbf{x} \wedge dr + 4 Vol_5\\
G_3 &= \delta G_3^{(1)} \\
\end{split}
\eeq
with
\beq
\delta G_3^{(1)} = dx^+ \wedge W
\eeq

We assign $W$ weight one under a Lorentz boost. The only term that can be generated at subsequent orders is the Schr\"odinger term in the metric,
\beq
\delta g_{AB}^{(2)} = g_{++}\delta_{A+}\delta_{B+}
\eeq
This fact ensures that the $\partial_{-}$ Killing vector stays null at all orders. Note that we have merely used the Lorentz symmetry of AdS, so this same logic applies to null perturbations of any Lorentz-invariant background by a massive supergravity field. In particular, this argument makes no use of supersymmetry or scale invariance of the near-horizon limit. Note also that the resulting Schr\"odinger solutions will be at zero temperature, and of the same curvature scale as their AdS counterparts: $L_{Sch}=L_{AdS}$.

This procedure can also be carried out by seeding the perturbation with the mixed KK vectors ($C$), or with the massive spin-2 field ($h$). These first-order choices just preserve different symmetries. In the former case, identical arguments to the above truncate the perturbation theory at second order; in the latter case, since our initial perturbation is of weight two, the linearized solution is in fact a full solution.

In fact, in this example the truncation of perturbation theory works even better than one might have expected.  As noted earlier, in the final equation in  (\ref{eqns1})  it would be consistent with the Lorentz symmetry to have a $|dC|^2$ term appear alongside $|W|^2$, but the coefficient of this term is apparently zero.  This may be a consequence of supersymmetry, and as we will see, has implications for the robustness of the $C \neq 0$ solutions against higher derivative corrections.

To actually obtain the Schr\"odinger metric, we turn now to the dilatation symmetry. The full nonlinear solution at hand is not Lorentz invariant, so it is consistent to allow for anisotropic scale invariance (\ref{scaling}).
This fixes the fields $W$ and $g_{++}$ to take their canonical Schr\"odinger scaling, and enforcement of the Bianchi identity on the three-form flux implies its closure. Parameterizing the metric as $g_{++}=r^2h$ gives
\beq\label{fields2}
\begin{split}
W &= d(r^z\sigma)\\
h &= r^{2z-2}q\\
\end{split}
\eeq
as in (\ref{fields}). Both $h$ and $W$ can have functional dependence only on $CY_3$ coordinates $(r,\Omega_5)$: specifically, $x^-$ is ruled out because of the $x^{\pm}$ scaling argument, and $\textbf{x}$ is ruled out to preserve rotational symmetry along the brane.

Using the dilatation invariance, general covariance, and the fact that $z=1$ must be an AdS solution for which we can set $W=0$, one can essentially reproduce the reduced field equations (\ref{fieldeqns}), including the $z$-dependence. Focusing on the spin-2 equation, we know that a two-derivative operator, covariant with respect to the $CY_3$ on which the fields are defined, must act on $h$. Scale invariance determines the form of the equation as
\beq
\nabla^2_{CY_3}(r^xh) = \alpha r^{x}|W|_{10}^2
\eeq
for some constants $x,\alpha$, and where the square of $W$ is taken with respect to the 10-dimensional metric. Each side has Lorentz weight two. Substituting from (\ref{fields2}) gives
\beq
\left(\nabla^2_5 + (2z+x-2)(2z+x+2)\right)q = \alpha \left(z^2|\sigma|^2_5 + |d\sigma|^2_5\right)
\eeq

We impose one final constraint, which is that when $z=1$, AdS is a solution with $\sigma=0$ and $q$ constant. This determines $x$ as
\beq
x(x+4)=0
\eeq
The choice $x=0, \alpha=-1$ is evidently made by plugging the ansatz into the Einstein equation itself. Up to this ambiguity, we have thus determined the relation between $z$ and the spin-2 mass, which was unknown \textit{a priori}, in contrast to the spectrum of $z$ in the well-studied massive vector model. \\

Thus, it is clear why the field equations (\ref{fieldeqns}) for the $\sigma=0$ solutions are simply those of the linearized level, and why the $\sigma \neq 0$ solutions have only a quadratic correction to those equations appearing in the Einstein equation. By using the \textit{spacetime} symmetries of the AdS vacuum we have essentially reproduced all aspects of the nonlinear solution in this perturbative context; we will see in the next section that we can also use the \textit{supersymmetry} of an AdS solution to make stronger statements about the existence of its Schr\"odinger deformations away from the supergravity approximation.

First, let us explain the existence of these Schr\"odinger fixed points from the CFT side.

\subsection{On the boundary: conformal perturbation theory}

Given the AdS/CFT correspondence, and its generalizations, it is instructive to give a version of
our arguments that applies to the boundary theory.  As we will see, this corresponds to a simple
exercise in conformal perturbation theory.
The response of a CFT to a null vector perturbation was studied  (along with other aspects of
holography for Schr\"odinger spacetimes) in \cite{Costa:2010cn,Guica:2010sw}, and we briefly
compare our conclusions at the end of this section.

Writing the CFT coordinates as $(x^\pm, \xb)$, where $x^\pm$ are
lightcone coordinates, the generators corresponding to dilatations and Lorentz boosts in $x^\pm$
act as
\beq\label{relgens}
\begin{split}
{\cal D}:\quad &x^+ \rar  {x^{+}}'={x^+ \over \lambda}\, ,\quad x^- \rar {x^-}'= {x^- \over \lambda}\, ,\quad \xb \rar \xb'= {\xb \over \lambda}  \\
{\cal L}:\quad  & x^+ \rar  {x^{+}}'={x^+ \over \kappa}\, ,\quad x^- \rar {x^-}'=\kappa  x^-\, ,\quad \xb \rar \xb'= {\xb }
\end{split}
\eeq

Now consider an operator $\cO_-$, whose dilatation weight is $\Delta$, and whose behavior under
Lorentz transformation is as indicated by its index structure.  Under a combined dilatation and Lorentz
transformation it transforms as
\beq
 \cO_-(x^+,x^-,\xb) \rar \cO_-({x^+}',{ x^-}',\xb') = \kappa^{-1}  \lambda^{\Delta}\cO_-(x^+,x^-,\xb)
\eeq
Adding this operator to the Lagrangian obviously breaks Lorentz invariance, and it also breaks
scale invariance if $\Delta \neq d$.    However, if we write $\Delta$ in the form
\beq\label{vect}
\Delta = d+z-1
\eeq
we see that under the non-relativistic dilatation generator, $\cD_z$,  defined as
\beq
\cD_{z} = \cD + (z-1)\cL
\eeq
and which acts as
\beq\label{nonrelgens}
\begin{split}
{\cal D}_{z}:\quad &x^+ \rar {x^{+}}'= {x^+ \over \lambda^z}\, ,\quad x^- \rar {x^{-}}'={x^- \over \lambda^{2-z}}\, ,\quad \xb \rar \xb'= {\xb \over \lambda}
\end{split}
\eeq
the operator $\cO_-$ acquires weight $\Delta_z=d$ under $\cD_z$, and so $ \int\! dx^+ dx^- d^{d-2} \xb \cO_-$
is scale invariant, as noted in  \cite{Costa:2010cn,Guica:2010sw}.

Similarly, a spin-2 operator $\cO_{--}$ will be marginal under $\cD_z$
provided that its relativistic conformal dimension is $\Delta = d+2(z-1)$.

Perturbing the action as
\beq
S \rar S + g_+  \int\! dx^+ dx^- d^{d-2} \xb ~\cO_- + g_{++}  \int\! dx^+ dx^- d^{d-2} \xb ~\cO_{--}
\eeq
thus preserves, to  first order in the couplings $(g_+, g_{++})$,   scale invariance generated by $\cD_z$.

In parallel to our discussion on the gravity side, we can consider the couplings $(g_+, g_{++})$ as  ``seeds",
and study the renormalization group beyond first order to see whether scale invariance survives.
We proceed using conformal perturbation theory (see \cite{Polchinski:1998rr}, section 15.8).   Inside the path integral
we expand $e^{-S}$ as a power series in the couplings.  Ultraviolet divergences can occur when two or more
operator insertions coincide in position space, and we regulate these by cutting out a small ball of
radius $\Lambda^{-1}$ around each operator, and adding counterterms to remove the divergences.   A breakdown of scale invariance is then signalled by the appearance of  $\log \Lambda$ terms, since
their removal introduces a scale $\mu$.

 At second order in perturbation theory a log
divergence can occur when two operators collide and produce a factor of $\int{dx^+ dx^- d^{d-2} \xb \over x^+ x^- |\xb|^{d-2}}$.
In particular, this will happen when the OPE of two vector operators behaves as
\beq
\cO_-(x^+,x^-,\xb) \cO_-(0,0,{\bf 0}) \sim {C\over x^+ x^-|\xb|^{d-2}} \cO_{--}(0,0,{\bf 0}) + \cdots
\eeq
where $\cO_{--}$ is a marginal (with respect to $\cD_z$) spin-2 operator.
In order to renormalize the theory we need to add to the action the operator $\cO_{--}$ with
a coefficient that depends logarithmically on scale.  The resulting theory thus breaks scale invariance
at second order in $g_+$ unless $C=0$, with a nonzero $\beta$-function $\beta_{++} \sim C (g_+)^2$.

This behavior corresponds to what we see on the gravity side.    Consider an AdS vacuum whose
field content includes a massive vector and a massive spin-2 field, whose masses are such that each
admits a linearized solution preserving  invariance under scale transformation with exponent $z$.
In particular, consider the system of equations (\ref{eqns1}), where we assume that the
appropriate CY$_3$ harmonics exist. Now seed a solution with the massive vector $W$.
According to the last equation in   (\ref{eqns1}) this will source $h$ at quadratic order in $W$.
By assumption, at linear order $h$ admits a solution $h \sim r^{2z-2}$, in order to be compatible
with the non-relativistic scale invariance. However, at quadratic order in $W$ we will have
to shift the power law of $h$ in order to solve the field equations.     Expanding in $W$,
this gives rise to a term $h \sim |W|^2r^{2z-2}  \ln r$, whose presence is the bulk analog of the logarithmically running coupling in the CFT.
Scale invariance  survives only if the source for the marginal $h$ mode
vanishes, which is the bulk analog of the CFT condition $C=0$.

In the CFT, checking scale invariance at orders $g_+ g_{++}$ and
$g_{++}^2$ means looking for marginal spin-3 and spin-4 operators appearing in the OPEs $\cO_- \cO_{--}$ and $\cO_{--} \cO_{--}$.  Nonzero OPE coefficients lead to nonzero $\beta$-functions for the spin-3 and spin-4 couplings.   Going beyond second order in perturbation theory,  we will similarly need to check for the appearance of operators of ever higher spin in OPEs.   How do we deal with this?
Here the key point is that in the supergravity limit the bulk theory contains no fields of spin larger than
two.  Such modes would correspond to stringy excitations, which we can think of as having been integrated
out, yielding $\alpha'$ corrections in the supergravity action.     Similarly, in order for the CFT
to have a dual supergravity description it is necessary that all single trace operators of higher spin should acquire
scaling dimensions that are parametrically large in the large $(\lambda, N)$ limit. So for such CFTs, there will not exist any single trace marginal operators beyond spin-2.\footnote{This argument does not exclude the possible
appearance of   multi-trace marginal operators of spin $>2$.   In the bulk, the presence of multi-trace operators shows up by modifying the boundary conditions  \cite{Berkooz:2002ug,Witten:2001ua} for fluctuations, but this has no effect on the vacuum solution itself in the classical, large $N$ limit.  If we restrict attention to the vacuum solution, then on the CFT side we can pretend that the multi-trace operators do not exist, and this is what we do in the following.   The effect of such multi-trace operators, if present, would show up in the computation of correlation functions, but this remains to be worked out.}    Therefore, the full set of renormalization group
equations governing the flows seeded by $(g_+, g_{++})$ consists
of the equations $\beta_+=0$, $\beta_{++} \sim C (g_+)^2$.
Existence of the fixed point thus boils down to checking the single condition $C=0$. As noted above, this matches the behavior seen in the bulk.

The vanishing of $C$ will follow from symmetry in some theories.
Namely, if the theory has a global symmetry group under which the
marginal spin-1 and spin-2 operators transform, $C$ will vanish unless
the product of two spin-1 representations contains the spin-2 representation.   For example, in the case of ${\cal N}=4$ SYM, operators will transform in
representations of the $SU(4)$ R-symmetry. The same considerations
apply on the gravity side, where for example in (\ref{fieldeqns}) the source terms appearing
on the right hand side are subject to the rules governing products of harmonics. (As noted earlier, when $SE_5 = S^5$, $C$ vanishes for all $z$.)

In the bulk, solving the last equation in   (\ref{eqns1}) will involve
turning on a tower of massive spin-2 two modes with amplitudes proportional to the square of the vector field.  We can make
a parallel observation on the CFT side.   Let us consider a
tower of generically \textit{non-marginal}  spin-2 operators $\cO_{--}^{(n)}$. In order to talk about the flow of their couplings $g^{(n)}_{++}$, let us now work in terms of the Wilsonian renormalization group, where we keep track of all couplings, not just the marginal and relevant ones.  Lorentz invariance now constrains their $\beta$-functions to be of the form
\beq\label{232}
\begin{split}
\beta_{++}^{(n)} & = \left( \Delta_z[\cO_{--}^{(n)}] -d \right) g_{++}^{(n)}+   C^{(n)} ( g_+)^2
\end{split}
\eeq
where $ \Delta_z[\cO_{--}^{(n)}]$ are the operator dimensions at the
Lorentz invariant fixed point.   We are including only the effect of
couplings with purely $+$-type indices, since all other couplings can be consistently set to zero by Lorentz invariance.  Along
with the fact that the equation $\beta_+=0$ is uncorrected by the presence of $g_{++}^{(n)}$, which again follows from Lorentz invariance,  we find the fixed point by taking
\beq
\label{betasol}
g_{++}^{(n)} = - {C^{(n)} \over \Delta_z[\cO_{--}^{(n)}] -d } (g_+)^2
\eeq
We can now see the parallel with the gravity side.  According to (\ref{fieldeqns}),  scale
invariant solutions are found by solving an inhomogeneous Laplace equation for $q$, with the
source given by the square of the vector field.    The solution can be decomposed into the
harmonics for the massive spin-2 fields.  The solution will then contain a tower of massive spin-2 fields
with amplitudes proportional to the square of the vector, which is what we have in  (\ref{betasol}).   To make this connection precise
we would like a better understanding of the bulk interpretation of
the Wilsonian couplings;  see \cite{Heemskerk:2010hk,Faulkner:2010jy} for recent work in this direction.

Finally, let us remark on references \cite{Costa:2010cn,Guica:2010sw}, which argued that couplings for vector operators $\cO_-$
are exactly marginal with respect to non-relativistic scale transformations.    The arguments in  \cite{Costa:2010cn,Guica:2010sw}
were based on showing that the 2-point function for $\cO_-$ is uncorrected by the addition of
$\cO_-$ to the Lagrangian.   This establishes that $\beta_+ =0$; however one also needs to verify
that $\beta$-functions for other couplings vanish as well.   As we discussed above,  this corresponds to checking that no marginal spin-2
operators appear in the OPE of the vector operators, and this has a
direct correspondence with the field equations in the bulk.

\subsection{A corollary: consistent truncations with massive KK modes}

Before moving on, we can apply our conclusion to the construction of consistent truncations of string/M-theory. By now, there is a large amount of technology for constructing consistent truncations with massive modes, e.g. \cite{mald, Gauntlett:2009zw, Cassani:2010uw, Liu:2010sa, Gauntlett:2010vu, Bena:2010pr, Cassani:2010na, O'Colgain:2009yd, Skenderis:2010vz}, and it was in this manner that the first string/M-theory embeddings of Schr\"odinger solutions were found in \cite{mald}. We have shown here that one can always elevate the linearized vector and spin-2 fluctuations around AdS to part of a full, nonlinear Schr\"odinger solution; but if one can elevate some of these fields to the nonlinear level on the level of the action itself, then it is clear that Schr\"odinger solutions to such theories always exist.

Therefore, a corollary to our argument is that \textit{anytime there exists a consistent truncation of an AdS$ \, \times \, \M$ KK spectrum that includes massive vector and/or spin-2 fields, the truncated theory admits the associated Schr\"odinger solutions.}

When can this be done? Although some recent work \cite{Gauntlett:2009zw, Gauntlett:2010vu} has argued for the possibility of including massive spin-2 fields in truncations to maximally supersymmetric supergravities, old lore \cite{stelle} and new evidence \cite{liupope} point to the contrary. If it is indeed true that no spin-2 fields can be so included -- supersymetrically or otherwise -- then the Schr\"odinger solutions built on diagonal vectors that also turn on a tower of spin-2 fields cannot be found as solutions of consistent truncations. In type IIB solutions, this excludes all but those that can be attained by a TsT transformation.

On the other hand, the solutions built from mixed vectors have no associated spin-2 fields. In principle, all of these solutions are ripe for embedding; in practice, all massive truncations to date have only included vector harmonics that sit at the base of their respective KK towers.

It was conjectured in \cite{Liu:2010sa} that a necessary condition for the consistency of the inclusion of a massive KK mode is that the field sits at the bottom of its tower, based on consideration of structure groups of compactification manifolds. If we take this to be true as well, then the possibilities for embedding Schr\"odinger solutions into string/M-theory by fitting them into consistent truncations are severely limited; for example, no more of the D3-deformed Schr\"odinger solutions can find such an embedding.\footnote{Later, we will present new solutions with Sch$_7$ symmetry based on M5 branes; according to these criteria, only one of them could be a solution to a consistent truncation on AdS$_7 \times S^4$.}

The flipside of this would be the implication that \textit{anytime} a massive vector is included in a consistent truncation, it can be used to construct a Schr\"odinger solution. This would provide a trivial prescription for identifying non-relativistic vacua, and at least gives a rule of thumb. We give an example later on of a supersymmetric truncation to a $d=4$, $\N=2$ gauged supergravity arising from wrapped M5 branes that includes a massive vector multiplet \cite{wrappedm5}, and show that, indeed, two Schr\"odinger solutions exist in correspondence to the theory's two AdS$_4$ vacua.

\section{Quantum/string corrections}

We turn now to the study of  corrections to Schr\"odinger solutions due to quantum and string effects, which hinges on analyzing  corrected KK spectra of AdS vacua. In what follows, there is no distinction made between $\alpha'$ corrections and $G_N$ corrections, as both enter on equal footing as higher derivative terms.   For convenience we sometimes refer to these collectively as $\alpha'$ corrections.

Let us first ask whether the existence of a solution is affected by $\alpha'$ corrections.

For Schr\"odinger solutions built on KK vectors and gravitons that do not couple to each other 
 -- these are the $q \neq 0$ and $\beta \neq 0$ solutions of section 2.1, for example -- we can quickly establish that \textit{there will always exist Schr\"odinger solutions at every order in a higher derivative expansion}. 
Assuming that the curvature does not become so large as to invalidate the gravity description, then there will always exist Schr\"odinger solutions made by utilizing the KK spectrum around the new, corrected vacuum in the preceding manner. This will generically induce a shift in $z$ and $L_{Sch}$, and is a generalization of a statement made in \cite{adams}. Using the dual arguments on the CFT side made in the previous section, the lone effect of heavy operators is to rescale dimensions such that a given operator is now marginal with respect to the new, shifted non-relativistic dilatations.

On the other hand, for Schr\"odinger solutions built on KK modes that \textit{do} couple, i.e. when the graviton mode obeys an \textit{inhomogeneous} Laplace equation -- these are the $q,\sigma \neq 0$ solutions of section 2.1, for example -- there may not be a solution upon including $\alpha '$ corrections. $z$ is determined by the vector spectrum, which will be shifted by the correction terms, and this induces a change in the quadratic source term: consequently, the corrected Einstein equation may not be soluble for a fixed vector harmonic. On the CFT side, the statement is that with respect to the new non-relativistic dilatations, there may exist a marginal spin-2 operator that appears in the OPE of two marginal vectors. In other words, the first term in \eqref{232} may vanish, eliminating the fixed point.


What is required of the AdS background for these corrections to vanish? As one might expect, the answer is supersymmetry. And as we show momentarily, certain families of Schr\"odinger solutions built from perturbations of the maximally supersymmetry AdS spacetimes are exact.\footnote{In what follows, we use the term ``exact'' to mean that a solution is unrenormalized to all orders in quantum and string corrections to the tree-level supergravity action. The existence of bonafide string-scale Schr\"odinger solutions is a matter which we have not investigated here.}

The outline for this section is as follows. In 3.1, we utilize supersymmetry in presenting the conditions for nonrenormalization, and the aforementioned exact solutions. In 3.2, we explicitly show how one class of exact solutions with Sch$_5$ symmetry, obtained by deformation of AdS$_5 \times S^5$, remains a solution to $O(\alpha'^3)$ despite its reduced isometry and supersymmetry. This is done by direct calculation with the conjectured metric and five-form flux correction terms at this order in the type IIB supergravity action.

\subsection{Uncorrected Schr\"odinger solutions from short multiplets}

In the presence of supersymmetry, there will typically be some degree of nonrenormalization of the AdS background and its KK spectrum. In cases for which the AdS background itself is robust against renormalization to some order in a derivative expansion, Schr\"odinger solutions built from perturbations of this background are similarly robust, as $L_{AdS} = L_{Sch}$.

The most useful fact for our purposes is basic: if a supergravity mode lies in a shortened multiplet of the corresponding AdS supergroup, then the exponent appearing in a power law profile for such a mode cannot get renormalized at any order in higher derivative corrections.    This can be understood intuitively as follows.   The action of the  dilatation operator directly relates the exponent to the scaling dimension of the dual boundary CFT operator; but operators in short multiplets have protected scaling dimensions.
Since $z$ appears as such an exponent, it will hence  not be renormalized. Shortened multiplets occur, for instance, for \textit{all} supergravity modes in maximally symmetric spacetimes because a long multiplet necessarily has fields with spins greater than two. Such shortened multiplets are actually ubiquitous in KK spectra, as many supergroups without maximal symmetry possess some number of short multiplets.

This immediately implies that if an AdS supergroup admits shortened multiplets, the Schr\"odinger solutions built from the massive fields in those multiplets will receive some level of protection from renormalization to all orders. Specifically, any property of the solution that depends only on the mass of such a field will be unrenormalized. Of course, this is the case for the dynamical exponent $z$ which is determined by either a vector or spin-2 equation; see (\ref{fieldeqns}). Therefore, we conclude that \textit{if the KK field that determines $z$ sits in a short multiplet, $z$ will remain uncorrected to all orders in an $\alpha'$ expansion.}

The reason that all solutions are not \textit{fully} protected is that the perturbation theory around AdS truncates at quadratic, not linear, order. In our canonical D3 brane example of section 2.1, those solutions formed from a diagonal vector ($\sigma \neq 0$) obey a spin-2 field equation corrected by a quadratic term. Thus, even if the spin-2 field $h$ is in a short multiplet, the solution will receive a correction. On the other hand, the solutions built on linear Laplace equations receive no renormalization at all, because \textit{all} properties of the solution are derived from field equations which remain uncorrected in the full solution.

We summarize these results as follows:\\

Suppose we have an AdS$\times \mathcal{M}$ solution which remains unrenormalized to $n^{th}$ order in $\alpha'$ or $l_P$. This implies that $L_{Sch}$ will follow suit. We now ask what other renormalization can occur. Suppose the KK spectrum contains both vectors and gravitons in short multiplets, and we turn these fields on to generate a Schr\"odinger geometry. The following nonrenormalization theorems will hold:
\begin{itemize}
\item \textbf{Vector and graviton do not couple}: The reduced field equations are simply those of the linearized fluctuations. So the full solution, modulo $L_{Sch}$, remains unrenormalized to all orders in higher derivative corrections.

\item \textbf{Vector and graviton couple}: This requires a tower of nonzero gravitons as well, which collectively solve a Laplace equation with a quadratic source. This solution will have $z$ unrenormalized, but the functional dependence of $g_{++}$ on $\mathcal{M}$ coordinates will be corrected at higher orders. (Again, $L_{Sch}$ is unchanged to $n^{th}$ order only.)
\end{itemize}

All of these arguments can be straightforwardly made on the CFT side, in accordance with the discussion in subsection 2.3.

In application to our D3-brane example of section 2, note that the uncoupled solutions (mixed vector and spin-2) are supersymmetric, generically preserving one half of the Poincar\'e supersymmetries of the AdS background \cite{hartnoll, gaunt2}. The coupled solutions (diagonal vector) can be made supersymmetric, but need not be \cite{gaunt, gaunt3, bobev}.

\subsubsection{Exact solutions}
When the AdS spacetime around which we perturb is maximally supersymmetric, $L_{AdS} = L_{Sch}$ receives no rescaling to all orders in higher-derivative corrections \cite{raj}, and all supergravity fields belong to short multiplets. Such spacetimes are the near-horizon geometries of the conformal branes in type IIB string theory and M-theory. Therefore, \textit{the towers of Schr\"odinger solutions based only on mixed vector and spin-2 perturbations of the maximally supersymmetric AdS$_{4/5/7} \times S^{7/5/4}$ vacua are exact.} These solutions preserve eight Poincar\'e supercharges \cite{gaunt}, and are of course not maximally isometric. As such, we add them to the short list of exact solutions with less than maximal supersymmetry and isometry, including the non-supersymmetric plane wave spacetimes studied in \cite{horowitz} and the 1/4 BPS AdS$_2\times S^2$ vacuum of $\N=2, d=4$ supergravity without matter multiplets \cite{raj}.

On the CFT side, this says that there exist Galilean-invariant vacua for arbitrary 't Hooft coupling and number of colors. In $\N=4$ SYM, the $(2,0)$ theory on M5 branes and the CFT on M2 branes, the conformal phase can be broken to a Galilean-symmetric phase with $z$ fixed as a function of $\lambda$ or $N$.\\

When the spheres are replaced by manifolds with less isometry, the resulting AdS vacua will have partial or no supersymmetry. Still, it is known in some cases that $L_{AdS}$ is norenormalized to a certain order in the derivative expansion. For example, using the conjectured $O(\alpha'^3)$ correction terms of type IIB involving the metric and five-form only \cite{paulos, greenstahn}, it has been shown that the AdS$_5 \times SE_5$ vacuum is unrenormalized to this order. Thus, the solutions with Sch$_5$ symmetry built by perturbing this background with massive fields in short multiplets, while keeping $\sigma=0$, are unrenormalized to this order too. The same goes for the AdS$_3 \times S^3 \times M_4$ geometry of the strongly coupled D1-D5 system, and its associated Sch$_3$ solutions which we present in section 4.

Indeed, the $SU(2,2|1)$ supergroup of the AdS$_5 \times SE_5$ vacuum is one of many that contains short multiplets. The shortening conditions for $SU(2,2|\N)$, with $\N=1,2,4$, are summarized in \cite{ferrara}. The most well-studied example of a geometry dual to an $\N=1$ SCFT, AdS$_5 \times T^{1,1}$, has both spin-2 and spin-1 fields in semi-long multiplets \cite{ferrara2}, dual to boundary operators of protected, rational conformal dimension. So solutions based on these fields are unrenormalized to at least $O(\alpha'^3)$.

Other AdS vacua whose KK spectra contain states in short multiplets include:  AdS$_3 \times S^3$ solutions of various $d=6$ supergravities, the multiplet structure of which was given in \cite{deboer}; AdS$_4 \times SE_7$ solutions, the supergravity states of which fall into unitary irreducible representations of $OSp(2|4)$ (e.g. \cite{Ceresole:1984hr, Fabbri:1999mk}); and orbifolded AdS vacua.

\subsection{At $O(\alpha'^3)$, explicit nonrenormalization of Sch$_5$}
We focus on type IIB Sch$_5$  solutions because, relatively speaking, we know a lot about type IIB correction terms. We possess an explicit form for the terms involving only the metric and five-form up to $O(\alpha'^3)$ relative to the supergravity action; we know that the supersymmetry of AdS$_5 \times SE_5$ vacua renders their length scales fixed to this order; and we know that when $SE_5 = S^5$, the solution is maximally symmetric, maximally supersymmetric, and hence unrenormalized to all orders, inclusive of corrections from \textit{all} type IIB fields.

The full set of metric and five-form corrections at $O(\alpha'^3)$ was presented in \cite{paulos}, building on the conjecture of \cite{greenstahn}. There exists a scheme, obtained by appropriate definition of the metric, in which the metric dependence can be written in terms of only the Weyl tensor, $C_{ABCD}$. All appearance of the five-form flux is via the following two-derivative tensor \cite{deHaro:2002vk},
\beq\label{T}
\T_{ABCDEF} = i\nabla_AF_{BCDEF} + \frac{1}{16}(F_{ABCMN}F_{DEF}^{\phantom{DEF}MN}-3F_{ABFMN}F_{DEC}^{\phantom{DEC}MN})
\eeq
where $[A,B,C]$ and $[D,E,F]$ should be independently antisymmetrized, and the two sets symmetrized under interchange. Schematically, the corrections to the Lagrangian take the form
\beq\label{terms}
C^4 \, , \quad C^3 \T \, , \quad C^2 \T^2\, , \quad C \T^3 \, , \quad \T^4
\eeq
along with their complex conjugates, all entering at $O(\alpha'^3)$; the full set of contractions is determined in \cite{paulos}. The terms enter the Einstein frame action as \cite{paulos, greenstahn}
\beq
S^{(3)} \sim \alpha'^3\int d^{10}x\sqrt{-g}f^{(0,0)}(\tau,\bar{\tau})(C^4+\ldots)
\eeq
where
\beq
f^{(0,0)}(\tau,\bar{\tau}) = \sum_{(m,n)\neq(0,0)}\frac{\tau_2^{3/2}}{|m+n\tau|^{3/2}}
\eeq
is a modular form \cite{gutperle} written in terms of the axio-dilaton $\tau = \tau_1 + i\tau_2 = \chi+ie^{-\phi}$, and the well-known $C^4$ term \cite{gross} is
\beq
C^4 = C^{ABCD}C_{EBCF}C_{A}^{\phantom{A}GHE}C^{F}_{\phantom{F}GHD} + \frac{1}{2}C^{ABCD}C_{EFCD}C_{A}^{\phantom{A}GHE}C^{F}_{\phantom{F}GHB}
\eeq
Beyond this order in $\alpha'$, the structure of the corrections is not known\footnote{It has not been proven that this is the full set of correction terms, though there is a great amount of evidence in favor.}.

For the AdS$_5 \times S^5$ background, in which $L_{AdS_5} = L_{S^5}$, superspace arguments have been used to prove that the solution is exact \cite{raj} . This is clearly satisfied at $O(\alpha'^3)$, as one can show that $C=\T=0$ on this solution. The uncorrected KK spectrum of energies is also evidently unrenormalized at this order: the corrections are quartic in $C$ and $\T$, so variation will never give a term linear in only one of these tensors.

Using this apparatus, we now consider corrections to the Sch$_5$ solutions obtained by deformations of AdS$_5 \times S^5$. We will explicitly demonstrate what we argued for earlier: that the metric and five-form solutions ($\sigma=0$) are unrenormalized, but that those with three-form flux ($\sigma \neq 0$) can receive a renormalization of the $S^5$ part of $g_{++}$. We consider these in turn, keeping $\beta=0$ for simplicity.

Before we begin, let us give away the punchline. It may seem like something of a mystery that these Sch$_5$ geometries, despite their greatly reduced supersymmetry and isometry, are as robust against these same corrections as the maximally supersymmetric and isometric AdS$_5 \times S^5$. As with many other aspects of Schr\"odinger holography, the null Killing vector is the key.

\subsubsection{Sch$_5$ solutions from AdS$_5 \times S^5$: $\sigma=0$}

This solution has metric and five-form flux only, so the only terms that can contribute to renormalization are the terms given above, plus terms linear in the axio-dilaton and complex three-form flux which can turn on a decoupled field. We delay treatment of these latter terms temporarily, as we turn to the $O(\alpha'^3)$ metric and five-form terms.

Our strategy will be to look for terms which contribute to $C$ and $\T$ -- each of which vanishes in AdS$_5$ -- and show that there are not enough terms to give nonzero contractions when plugged into the correction terms from (\ref{terms}).

The Sch$_5$ metric
\beq
ds^2 = r^{2z}q(dx^+)^2 + ds^2(AdS_5 \times S^5)
\eeq
corrects the vanishing AdS$_5 \times S^5$ Weyl tensor only by terms which have two lower + indices and no - indices. Specificially, nonvanishing components are
\beq\label{weyl}
C_{+r+\theta_i}  \, ; \quad C_{+\theta_i+\theta_j} \, ; \quad C_{+r+r} \, ; \quad C_{+\textbf{x}+\textbf{x}}
\eeq
where $\lbrace \theta_i\rbrace$ parameterize $S^5$. There are no lower - indices. Already, this implies that variation of the $C^4$ term will vanish on-shell, because the + indices have nothing to contract with: $g^{++}=0$.\footnote{Note that this same phenomenon for the Sch$_3 \times S^3$ solution from the D1-D5 system, to be presented in the next section, implies that the $C^4$ term does not renormalize that solution either \cite{gubser}.}

The correction terms are quadratic in $C$ and $\T$, so variation of these terms with respect to either $F_5$ or the metric will leave behind terms at least cubic in $C$ and $\T$. Because $g^{++}=0$, this implies that if $\T$ has no components with lower - indices, all variation of the correction terms will vanish: the lower + indices of either tensor will not have any upper + indices to contract with, and the quartic order guarantees enough index contractions to force such a contraction. So, we examine the structure of $\T$, given in (\ref{T}).

Recall that $\T=0$ in AdS$_5 \times S^5$. We can uncover the structure of this term in the Schr\"odinger vacuum without actually plugging in for the flux -- which is the same in both Schr\"odinger and AdS backgrounds, namely the sum of volume forms -- by asking what new components of the Christoffel connection and contravariant metric are introduced by the deformation.

The new components of the connection are
\beq
\begin{split}
&\Gamma^{r}_{++} = -qzr^{2z+1} \, , \quad \Gamma^-_{+r} = qr^{2z-3}(z-1) \, , \\
&\Gamma^{-}_{+\theta_i} = \frac{1}{2}r^{2z-2}\frac{\partial q}{\partial{\theta_i}} \, , \quad \Gamma^{\theta_i}_{++} = -\frac{1}{2}r^{2z}\frac{\partial q}{\partial{\theta_i}}  \\
\end{split}
\eeq
This implies that the flux remains covariantly constant in the Schr\"odinger background,
\beq
\nabla_AF_{BCDEF}=0
\eeq
This relies on the fact that the flux is merely the sum of the two volume forms, just as in AdS$_5 \times S^5$, and the fact that none of the new Christoffel components has an identical upper and lower index.

Now we examine the terms quadratic in $F_5$. Because the components of $F_5$ are the same as in AdS$_5$, the only changes will come from new contravariant components of the metric, of which there is one: $g^{--} \neq 0$. This implies the existence of a single new nonzero term. Consider the contraction
\beq
F_{ABC-G}F_{DEF}^{\phantom{DEF}-G} = F_{ABC-G}F_{DEF+}^{\phantom{DEF+}G} g^{+-} + F_{ABC-G}F_{DEF-}^{\phantom{DEF-}G} g^{--}
\eeq
The second term is the new term; the first term merely contributes to the condition $\T=0$ in the unperturbed AdS background. Looking at the new term, we see that $G \neq +$, otherwise it vanishes. So one index within each triplet $[A,B,C]$ and $[D,E,F]$ must be the + index. Hence, the new term takes the schematic form
\beq
\T_{+BC+EF} \sim F_{+BC-G}F_{+EF-}^{\phantom{+EF-}G} g^{--}
\eeq
where $B,C,E,F \neq -$. More specifically, the only nonvanishing terms are
\beq
 \T_{+xy+xy} \, , \quad \T_{+rx+rx} \, , \quad \T_{+ry+ry}
\eeq
because $F_{+-rxy} \neq 0$ is the only nonvanishing flux component with legs along $+$.

To summarize, the only nonvanishing components of $\T$ and $C$ have no lower - indices, and hence any variation of the correction terms will vanish on-shell.

This is a perhaps surprising  result. To further drive the point home, we point out an argument in \cite{paulos} that any solution with all fields trivial except for the metric and five-form that is more than 1/4 BPS is uncorrected at $O(\alpha'^3)$ by the entire set of terms above. Here, we have shown the same for these 1/4 BPS solutions. \\

We now turn to possible  terms linear in the axio-dilaton and three-form flux, which would act
as sources for these fields.   Symmetry under $G_3 \rightarrow -G_3$ rules out terms linear in $G_3$.
Terms linear in the axio-dilaton would, at this order, be multiplied against a scalar, eight-derivative
object constructed from the Weyl tensor and five-form field strength.   Further, any such scalars must
vanish when evaluated on the AdS background, since this is a  solution by assumption.   Schematically,
this would allow terms such as
\beq
 (\phi,\chi)\, C_{ABCD}\left((F_5)^6\right)^{ABCD}\, , \quad (\phi,\chi) \,\nabla_{A}C_{BCDE}\left((F_5)^5\right)^{ABCDE}\,  \ldots
\eeq
However, by examining the index structure and using the null Killing vector, one can check that all such
scalar terms vanish.  We conclude that the axio-dilaton is not sourced at this order.

\subsubsection{Sch$_5$ solutions from AdS$_5 \times S^5$: $\sigma \neq  0$}

Now we turn on three-form flux. This does not affect the metric and five-form correction analysis above. So if this solution is to be renormalized, it must come from the three-form terms. In the absence of knowledge of the actual form of the terms, the null constraint is not strong enough to rule this out.

Let us at least present the challenge. First, $F_5$ is not constrained to appear via the $\T$ tensor in these terms. This introduces a tensor component with a lower - index, $F_{+-rxy}$, as well as one without any + or - indices at all, $F_{\theta_1\ldots\theta_5}$. The metric still must appear via the Weyl tensor, still as in (\ref{weyl}), so the full set of nonzero components is
\beq
C_{+A+B} \, , \quad F_{+-rxy} \, , \quad F_{\theta_1\ldots\theta_5} \, , \quad G_{+r\theta_i} \, , \quad G_{+\theta_i\theta_j}
\eeq
Additionally, $G_3$ is not covariantly constant, so we can have $\nabla G_3$ appear in the correction terms.

Suppose we wish to ask what terms can contribute to the stress tensor.  The following term is explicitly nonvanishing on this background, and could arise at $O(\alpha'^3)$:
\beq
\begin{split}
(G^2F^6)_{++} &\sim G_{+BC}G_{+DE}F^{BCFGH}F^{DE}_{\phantom{DE}FGH}F^4\\ &\sim G_{+\theta_i\theta_j}G_{+\theta_k\theta_l}F^{\theta_i\theta_j\theta_m\theta_n\theta_p}F^{\theta_k\theta_l}_{\phantom{\theta_k\theta_l}\theta_m\theta_n\theta_p}F^4
\end{split}
\eeq
where $F^4$ includes all possible contractions. Another term is
\beq
\begin{split}
(G^2F^6)_{++} &\sim F_{+ABCD}F_{+E}^{\phantom{+E}BCD} G^{AGH}G^{E}_{\phantom{E}GH}F^4\\
&\sim F_{+-rxy}F_{+-}^{\phantom{+-}rxy} G^{-GH}G^{-}_{\phantom{-}GH}F^4\\
\end{split}
\eeq
The correct properties under symmetry are manifest for both of these terms.

Therefore, the null Killing vector does not imply an absence of renormalization of the $g_{++}$ component of these Sch$_5$ solutions. It is quite possible, if not likely, that the true three-form correction terms involve a tensor structure, analogous to $\T$ for the five-form and metric terms, that renders the corrections to the Sch$_5$ backgrounds zero.

\section{New Schr\"odinger solutions}

We put our theory into practice by presenting new Schr\"odinger solutions. Two of these are based on the compactification spectra of well-studied brane setups: the maximally supersymmetric M5 brane in flat space, and the half-supersymmetric D1-D5 system of type IIB with the D5 brane wrapped on $T^4$ or $K3$. The third construction presents evidence for our earlier suggestion that all consistent truncations of string/M-theory that retain massive modes admit Schr\"odinger solutions. In this instance, we build a pair of Sch$_4$ solutions dual to nonrelativistic phases of a 2+1 CFT living on M5 branes wrapped on special-Lagrangian 3-cycles of a Calabi-Yau.

The M5 construction includes two infinite families of solutions exact to all orders in $l_P$. The D1-D5 construction includes three families that are exact up to the order at which $L_{AdS_3}$ is rescaled.\footnote{It is known that at $O(\alpha'^3)$, $L_{AdS_3}$ is unrenormalized.}

\subsection{Sch$_7$ from M5 branes}
The KK spectrum around AdS$_7 \times M_4$ \cite{vanNieuwenhuizen:1984iz, Gunaydin2} contains two towers of mixed vectors, no massive diagonal vectors, and massive gravitons as always.  This spacetime has $L_{AdS_7} = 2L_{M_4}\equiv 2$. For $M_4 = S^4$, the upper branch of vectors has masses $m^2 = (l+3)(l+5)$ (where integer $l \geq 0$ henceforth), implying a (positive) $z$ spectrum of $z=6+2l$, in accordance with (\ref{kkd}). The massive gravitons have spectrum $m^2 = (l+1)(l+4)$, which implies a (positive) $z$ spectrum of $z=2+l$, in accordance with (\ref{kkgd}).

If one replaces $S^4$ with an $M_4$ which has nonzero second Betti number $b_2$, there will be an additional set of Yang-Mills gauge fields in $d=7$. These Betti vectors carry topological charge, descending directly from the three-form gauge field and appearing as a single harmonic level of diagonal vectors. Their masslessness gives an isolated Sch$_7$ solution with $z=-4$ and degeneracy $b_2$; the possibility $z=0$ is ruled out, as the Schr\"odinger vector in this case would be pure gauge.

We will realize each one of these solutions in $D=11$, again starting from the full M5-brane metric and then moving into the near-horizon. The most general solution\footnote{Conventions for $D=11$ supergravity are as in \cite{pakis}.} is
\beq
\begin{split}
ds^2 & = \Phi^{-1/3}(2dx^+ dx^- +h (dx^+)^2 +2 dx^+ C+d\textbf{x}^2) +\Phi^{2/3} ds^2(X_5) \\
G& = \star_{11}(dx^+ \wedge dx^- \wedge d\textbf{x} \wedge d\Phi^{-1}) +dx^+ \wedge V - \star_{11}(dx^+ \wedge d\textbf{x} \wedge d(\Phi^{-1} C)) \\
\end{split}
\eeq
This ansatz preserves rotational invariance. $C$ is a one-form, $V$ is a three-form, and $h$ is a function, all defined on the space $X_5$. The Maxwell and Einstein equations, and the Bianchi identity, imply
\beq
\begin{split}
&d\star_{X_5}\Phi = 0 \\
&d\star_{X_5}dC = 0\\
&d\star_{X_5}\left(\frac{V}{\Phi}\right)= 0\\
&dV = 0 \\
&\nabla^2_{X_5}h = -\frac{1}{\Phi} |V|^2_{X_5}\\
\end{split}
\eeq
where $|V|^2_{X_5} = \frac{1}{3!}V_{abc}V^{abc}$ with indices raised by the metric on $X_5$. Notice the extra factors of $\Phi$ in the $V$ terms relative to the D3 brane case (cf. equations (\ref{eqns1})).

Set $V=0$ for now. Zooming into the near-horizon region and writing $X_5$ as a cone over $M_4$,
\beq
ds^2(X_5) = dr^2 + r^2ds^2(M_4)
\eeq
we make the scale-invariant assignments
\beq
\begin{split}
\Phi = r^{-3}\\
C=r^{\frac{z-2}{2}}\beta\\
h=r^{z-1}q\\
\end{split}
\eeq
where $q$ and $\beta$ are a function and one-form, respectively, on $M_4$. When $V=0$, this is a solution so long as the reduced field equations
\beq
\begin{split}
&\Delta_4\beta = \frac{z(z-2)}{4}\beta \quad \text{where} \quad d\star_4\beta=0\\
&\nabla^2_4q + (z-1)(z+2)q=0\\
\end{split}
\eeq
are satisfied, with $-\nabla^2_4$ and $\Delta_4$ as the Laplace operator of $M_4$ acting on functions and transverse one-forms, respectively. For a spherical base space $M_4=S^4$,
\beq
\Delta_4\beta = (l+2)(l+3)\beta\, , \quad \nabla^2_4q+l(l+3)q=0
\eeq
and we obtain the anticipated spectra of $z$ outlined earlier. Notice that these two solutions can be consistently superposed, because there exist simultaneous eigenfunctions of $\Delta_4$ and $-\nabla^2_4$ for all $z$.

Because all KK fields sit in short multiplets of the superalgebra $OSp(8|4)$ \cite{Gunaydin2}, these solutions are exact. By inspection of the structure of this solution, we expect it to preserve eight Poincar\'e supersymmetries, in analogy with the M2 and D3 brane solutions of \cite{gaunt}.

Turning to $V \neq 0$, the discussion at the start of this subsection implies the existence of a single solution with $z=-4$ corresponding to deformation by the $b_2$ topological vector fields. Indeed, one can show that the following configuration solves the field equations,
\beq
V = d(r^{-2}\tau)
\eeq
where $\tau$ is a harmonic two-form on $M_4$, $d\tau=d\star_{M_4}\tau=0$. So $M_4$ must have $b_2 \neq 0$, and scale invariance of $G$ implies $z=-4$ as predicted, since $V$ must scale as $r^{z/2}$. The associated Einstein equation is
\beq
\nabla^2_{4}q + 10q = -4|\tau|_{4}^2
\eeq

Perhaps unexpectedly, this is but one of an infinite family of diagonal solutions, despite the absence of diagonal vectors in the spectrum of KK fields on AdS$_7 \times M_4$. We present the general construction in appendix A, along with a Lif$_6(z=2)$ solution obtained in a similar manner.

\subsection{Sch$_3$ from D1-D5 and F1-NS5 branes}
These solutions come from deformations of AdS$_3 \times S^3 \times M_4$, with $M_4=T^4,K3$. The KK spectra \cite{sezgin} around these two solutions are identical in everything but multiplicities of fields, on account of the different number of tensor multiplets in the chiral and non-chiral $d=6$ supergravities. They are somewhat more involved than the examples presented so far because of the prior reduction on $M_4$, combined with the self-duality properties of the tensor fields. As a result, we relegate a full treatment to an appendix and only present the solutions here.

Once again, we expect these solutions to have the same supersymmetric structure as their D3 and M2-brane counterparts: the mixed solutions should preserve half of the Poincar\'e supersymmetry of the D1-D5 background, and the diagonal solutions can, but need not, be supersymmetric. We have not confirmed this explicitly, however.

The KK reduction on $S^3$ is done at the level of the $d=6$ supergravity, which contains five self-dual and $n_T$ anti-self-dual tensor fields that descend from the complex three-form and five-form fluxes, where $n_T$ is the number of antisymmetric tensor multiplets.\footnote{When $M_4=T^4$, K3, one has $n_T=5,21$, respectively.}
\subsubsection{Diagonal vectors}
There are two towers of diagonal vectors, each with mass $(mL)^2 = (l+2)^2$, implying a $z$ spectrum of $z^2=(l+2)^2$, where $l \in \mathbb{Z}$. One tower descends from the four self-dual tensor fields that do not source the D1-D5 background, and the other from the $n_T$ anti-self-dual tensor fields.

The ansatz is
\beq
\begin{split}
&ds^2 = r^{2z}q(dx^+)^2+ 2r^2dx^+dx^- + \frac{dr^2}{r^2} + ds^2(S^3) + ds^2(M_4)\\
&iG_3 = 2(1+\star_6)d\Omega_3 + \delta F_3 - i\delta H_3\\
&F_5 = \delta F_5\\
\end{split}
\eeq
where $q$ is a function defined on $S^3$, $d\Omega_3$ is the invariant volume form on $S^3$, and $\star_6$ is the Hodge star operator on the spacetime transverse to the $M_4$. We have set $L_{AdS_3} = L_{S^3} = 1$. When $q=\delta H_3 = \delta F_3 = \delta F_5=0$, this is the D1-D5 solution.

Because the $d=6$ tensor fields descend from both $G_3$ and $F_5$, we simply need to turn on the components of $G_3$ and $F_5$ that give rise to tensor fields with the right $d=6$ self-duality property. This means that we can turn on anti-self dual parts of $F_3, H_3$ and $F_5$, but only self-dual parts of $H_3$ and $F_5$. This also implies that without doing any extra work, the F1-NS5 system obtained from the D1-D5 system by S-duality can also be engineered to give Schr\"odinger solutions in exactly the same way: from the six-dimensional perspective, S-duality merely shuffles the background flux to a different one of the five self-dual tensor fields.

Without further ado, the solutions are:
\begin{itemize}
\item Anti-self-dual RR two-form charge:
\beq
\delta F_3 = (1-\star_6)dx^+ \wedge d(r^z \sigma)
\eeq

\item NS-NS two-form charge:
\beq
\delta H_3 = (1\pm\star_6)dx^+ \wedge d(r^z \sigma)
\eeq

\item RR four-form charge:
\beq
\delta F_5 = \frac{1+\star_{10}}{\sqrt{2}}\left[(1\pm\star_6)dx^+ \wedge d(r^z \sigma)\wedge \alpha\right]
\eeq
\end{itemize}

$\sigma$ is a one-form on $S^3$, $\alpha$ is a harmonic two-form on $M_4$ of norm $|\alpha|^2=1$ and definite $M_4$ self-duality property, and the reduced field equations are
\beq\label{d1d5eqns}
\begin{split}
&\Delta_3\sigma = z^2\sigma \quad \text{where}\quad d\star_3\sigma=0\\
&\nabla^2_3q + 4z(z-1)q = -2\left(z^2|\sigma|_3^2 + |d\sigma|_3^2\right)\\
\end{split}
\eeq
with $-\nabla^2_3$ and $\Delta_3$ as the Laplace operator of $S^3$ acting on functions and transverse one-forms, respectively. Note that the source terms in the Einstein equation are of the same form as those in the D3-brane case, given in (\ref{fieldeqns}). For future reference, let us denote this quantity
\beq
\Lambda(\sigma)\equiv z^2|\sigma|_{\mathcal{M}}^2 + |d\sigma|_{\mathcal{M}}^2
\eeq

On $S^3$, $\Delta_3\sigma = (l+2)^2\sigma$, where $l \in \mathbb{Z}$, so taking the positive branch gives
\beq
z=l+2
\eeq
as expected.

Note that a $z=2$ solution was constructed in \cite{oz} as a TsT-transformed D1-D5 system, by utilizing the Reeb Killing vector of $S^3$, and studied more recently in \cite{Banerjee:2011jb}. (For the construction of another $z=2$ solution in a somewhat different setting, see also \cite{Orlando:2010yh}.) That solution has nonzero $H_3$ charge, both self-dual and anti-self-dual. This is consistent with our result, as the lowest eigenvalue of $\Delta_3$ is obtained for $\sigma$ Killing; in fact, we see that it is actually redundant, because it turns on two different KK vectors with the same mass. The way this TsT solution fits into the ladder of solutions is qualitatively identical to the TsT D3 brane solution: in particular, $z=2$ is the lone value which can give a direct product metric Sch$_3 \times S^3 \times M_4$ because $\Lambda(\sigma)$ is not constant otherwise. (See the appendix for more discussion on this point.)
\subsubsection{Spin-2 fields}
It is clear from the diagonal vector solutions that if we turn the vector off, we will get a Sch$_3$ solution from a spin-2 field alone, where $z$ is now determined by the scalar Laplacian on $S^3$ as
\beq
4z(z-1) = l(l+2) \,
\eeq
Note the agreement with (\ref{kkgd}). Starting at the first nonzero harmonic $l=1$, we have solutions with
\beq
z= \frac{3}{2}\, , \, 2 \, , \, \frac{5}{2} \, , \ldots
\eeq

\subsubsection{Mixed vectors}
There are three towers of mixed vectors, each with different mass spectra. Here, we construct Sch$_3$ solutions from two of them. One has $m^2 = l^2$, leading to a  $z$ spectrum of $z^2 = l^2$. The other has  $m^2 = (l+4)^2$, leading to a $z$ spectrum of $z^2 = (l+4)^2$.

The solution is
\beq
\begin{split}
&ds^2 = 2(r^2dx^+dx^- +r^zdx^+\beta) +  \frac{dr^2}{r^2} + ds^2(S^3) + ds^2(M_4)\\
&iG_3 = 2(1+\star_6)d\Omega_3 +(1+\star_6)dx^+\wedge d(r^z\beta)-\frac{r^2}{2}(1+\star_6)dx^+\wedge d(r^{z-2}\beta) \\
\end{split}
\eeq
where $\beta$ is a one-form on $S^3$ and obeys the eigenvalue equation
\beq
\Delta_3\beta = (z-2)^2\beta \quad \text{where} \quad d\star_3\beta=0
\eeq
This gives $z=-l, \,l+4$, consistent with our expectation. As with other mixed vector constructions, each branch of the eigenvalue equation accesses one branch of vectors. The final term in the flux appears anomalous compared to other mixed vector solutions; as detailed in the appendix, this is a peculiarity of the dimensionality of these solutions.

\subsection{Sch$_4$ from wrapped M5 branes}
Finally, we construct two Sch$_4 \times H^3/\Gamma \times S^4$ solutions of $D=11$ supergravity. This is based on a consistent (bosonic) truncation to an $\N=2, d=4$ gauged supergravity with one vector multiplet and two hypermultiplets, as performed in \cite{wrappedm5}. The M5 branes are wrapped on special-Lagrangian 3-cycles of $CY_3$, giving a $d=3, \N=2$ CFT at low energies; the supersymmetry is preserved by virtue of the choice of cycle. With the consistent truncation in hand, two AdS$_4$ duals can be found directly within the $d=4$ supergravity. Here, we show that these $d=3$ CFTs also have non-relativistic phases by finding the dual Schr\"odinger geometries.

Most details of the truncation are unnecessary for our purposes; we refer the reader to \cite{wrappedm5}, and use their notation in what follows.

The reduction is done first from $D=11$ supergravity on $S^4$ to maximal seven-dimensional gauged supergravity, and then further on a three-manifold of constant curvature, $\Sigma_3=S^3, \mathbb{R}^3, H^3$ or their quotients. One can parameterize the choice of $\Sigma_3$ by the sign of the curvature, $l = \pm1,0$, where $S^3$ is taken to have $l=1$. The theory contains the following field content: the metric; two three-forms $h_3^{\alpha}$, where $\alpha=1,2$; one two-form $B_2$; two one-forms $A_1, C_1$; and nine scalars comprised of $\phi, \lambda, \beta, \theta_{\alpha}, \chi_{\alpha}$ and a symmetric matrix $\T_{\alpha\beta}$, which parameterizes a coset $SL(2,\mathbb{R})/SO(2)$. In addition to the parameter $l$ which appears in the $d=4$ field equations, there is a gauge parameter $g$ with mass dimension one that comes from the $d=7$ gauged supergravity potential.

We will not need most of these fields to construct the Sch$_4 \times H^3/\Gamma \times S^4$  solutions.

There are two known AdS$_4$ vacua of this theory, and the masses of the fields in each vacuum were calculated in \cite{wrappedm5}. Only the first is supersymmetric, and is given as
\beq
e^{\phi}=2^{-1/20} \, , \quad e^{\lambda}=2^{1/10} \, , \quad l=-1 \, , \quad (gL)^2 = \sqrt{2}
\eeq
with all other fields turned off. The vector fluctuations around this background can be diagonalized to give vectors with masses $(mL)^2 = 0,4$, with $C_1$ as the massive vector. One can show that there is a Schr\"odinger solution, with
\beq\label{z1}
z(z+1)=4 \quad \Rightarrow \quad z=-\frac{1}{2} + \frac{\sqrt{17}}{2} \approx 1.56
\eeq
in which the AdS fields and parameters above are unchanged, and we turn on the form-fields in the following manner:
\beq
\begin{split}
C_1 &= r^zdx^+\\
B_2 &= -\frac{1}{\sqrt{2}g}\star_4dC_1 \\
F_2 = dA_1 &= -\frac{3}{g}\,dC_1\\
\end{split}
\eeq
with $x^+$ a lightcone coordinate of AdS$_4$, and $z$ as in (\ref{z1}). Hence, the full Sch$_4 \times H^3/\Gamma \times S^4$ metric reads
\beq
ds^2 = -r^{2z}(dx^+)^2 + r^2(2dx^+dx^- + dx^2) + \frac{\sqrt{2}}{g^2}\frac{dr^2}{r^2} + ds^2(H^3/\Gamma \times S^4)
\eeq
The massive vector sits in a long vector multiplet of $OSp(2|4)$, and so this solution will be subject to quantum corrections.

The second, non-supersymmetric AdS$_4$ solution is given as
\beq
e^{\phi}=6^{-1/4}10^{1/5} \, , \quad e^{\lambda}=10^{1/10} \, , \quad l=-1 \, , \quad (gL)^2 = \frac{5\sqrt{2}}{3\sqrt{3}}
\eeq
with all other fields turned off. The vector fluctuations around this background can be diagonalized to give vectors with masses $(mL)^2 = 0,\frac{28}{5}$; again, $C_1$ is the massive vector. One can show that there is a solution, with
\beq\label{z2}
z(z+1)=\frac{28}{5} \quad \Rightarrow \quad z=-\frac{1}{2} + \frac{3}{10}\sqrt{65} \approx 1.92
\eeq
in which the AdS fields and parameters above are unchanged, and we turn on the form-fields in the following manner:
\beq
\begin{split}
C_1 &= r^zdx^+\\
B_2 &= -\frac{\sqrt{6}}{14g}\star_4dC_1 \\
F_2 = dA_1 &= -\frac{27}{7g}\,dC_1\\
\end{split}
\eeq
where $dx^+$ is again lightcone coordinate of AdS$_4$, and now $z$ is as in (\ref{z2}).

A useful way to see that these solutions exist is to consider the $d=4$ field equations, and show that they can be truncated to those of a massive vector theory with the appropriate mass and cosmological constant, \textit{provided} that the vector is null. The definitions of the parameters $(L,g,l,m^2)$ used in these solutions fall out of this procedure. For a brief exposition of this, see appendix C.

The considerations  in earlier sections  tell us that these are but two of an infinite tower of Schr\"odinger deformations of these AdS$_4$ solutions. The others are implicit in the compactification spectrum. We were able to embed this one in a consistent truncation because, presumably, the massive vector $C_1$ sits at the bottom of a tower of vectors on $H^3/\Gamma$. This is supported by the fact that our solutions contain direct product metrics.

\section{Discussion}
Understanding Schr\"odinger solutions as Kaluza-Klein deformations of AdS has revealed some larger truths. We saw that they are universal, existing in infinite number for any AdS vacuum, by virtue of the maximal isometry of AdS; they can be robust against quantum and string corrections, by virtue of possible supersymmetry of AdS; and in some cases, they are exact. In essence, we have shown that the AdS/CFT correspondence is not only suggestive of a generalization to non-relativistic, Galilean gauge-gravity duality, but rather, the latter is truly contained within the former.

We expect these ideas to lend themselves to applications to condensed matter systems. Just as the universality of AdS leads to general thermodynamic results for holographically dual gauge theories (e.g. the KSS bound), one might expect that the universality of Schr\"odinger spacetimes implies similar statements about non-relativistic fixed points. Some work has already been done to show that the KSS bound is saturated for certain values of $z$ (e.g. \cite{Herzog:2008wg, Adams:2008wt} studied $z=2$); the spirit of the present work suggests a universal extension of some kind.

Our analysis relies on the Lorentz and scaling symmetries of AdS, which precludes an extension of the perturbation theory method to finite temperature. On the other hand, any asymptotically AdS spacetime will, at infinity, possess the symmetries and KK spectrum required for the existence of Schr\"odinger deformations. It would be interesting to further investigate, in this KK framework, the extent to which asymptotically AdS black hole solutions can be universally deformed.

An aspect of this work that may lend itself to the study of RG flows with Schr\"odinger endpoints is the discovery of the role of relevant operators in generating Schr\"odinger backgrounds. We showed in section 2 that the generic Sch$(z)$ solutions in string and M-theory involve a tower of spin-2 fields which are non-marginal with respect to non-relativistic dilatations $\mathcal{D}_z$ in the AdS vacuum. Some of these fields are relevant: any spin-2 field which has relativistic scaling weight $\Delta = d+2(z'-1)$ under $\mathcal{D}$ is relevant with respect to $\mathcal{D}_z$ when $z' < z$. In fact, in circumstances where the field theory possesses non-Abelian internal symmetries they may all be relevant.\footnote{One example is the set of solutions obtained by diagonal vector deformations of AdS$_5 \times S^5$, in which the global $SU(4)_R$ symmetry leads to a restricted set of lower harmonics appearing in the solution \cite{gaunt, bobev}.} Supposing that one wants to construct an interpolation between a UV CFT and its IR Schr\"odinger phase of the same effective spacetime dimension, one may be able to utilize one of these relevant operators to stabilize an IR Schr\"odinger geometry.

Continuing with the topic of RG flows, it would be interesting to look for one that connects the Sch$_4 \times H^3/\Gamma \times S^4$ solution at low energies to an AdS$_7 \times S^4$ solution at high energies, in the spirit of \cite{maldnunez}. The picture in the bulk is of an  M5 brane, extended at asymptotic infinity, wrapping itself around $H^3/\Gamma$ and turning on flux at small radii, dual to a nontrivial RG flow across dimension in which the scale invariance of the theory becomes anisotropic in the infrared. An example of this sort of behavior was recently found in \cite{D'Hoker:2010ij} (though with Sch$_3$ symmetry and hence without Galilean boosts). It may also be that the Sch$_4$ solution built from a deformation of the supersymmetric AdS$_4$ solution preserves some of the eight supersymmetries, in which case one would be motivated to look for an analytic RG flow, generalizing the work of \cite{maldnunez} to the non-relativistic regime.

\bigskip \bigskip

\noindent
{\Large \bf Acknowledgments}\medskip

We thank Martin Ammon, Eric D'Hoker, Miguel Paulos, Brian Shieh and Phil Szepietowski for helpful conversations. E.P. is supported in part by a Leo P. Delsasso Fellowship from the UCLA Department of Physics and Astronomy.  P.K. is supported in part by NSF grant PHY-07-57702.

\noindent
\appendix
\section{More non-relativistic solutions from M5 branes}
\setcounter{equation}{0}
\subsection{Diagonal Sch$_7$ solutions}
Here, we present a family of Sch$_7$ solutions which do not appear to lie in correspondence with any KK fields on AdS$_7 \times M_4$.

Starting from the near-horizon ansatz
\beq\label{m5ansatz}
\begin{split}
ds^2 & = r(2dx^+ dx^- +h (dx^+)^2 +d\textbf{x}^2) +\frac{dr^2}{r^2} + ds^2(M_4) \\
G& = \star_{11}(3r^2dx^+ \wedge dx^- \wedge d\textbf{x} \wedge dr) +dx^+ \wedge V \\
\end{split}
\eeq
the field equations are
\beq
\begin{split}
&d\star_{X_5}\left(Vr^3\right) = 0\\
&dV = 0 \\
&\nabla^2_{X_5}h = -r^3|V|^2_{X_5}\\
\end{split}
\eeq
where as usual, $X_5$ is the cone over $M_4$. Scale invariance implies that $V$ scales as $r^{z/2}$.

Following the strategy of the cases in which we \textit{do} expect such diagonal solutions, we make the ansatz
\beq
V=d(r^{z/2}\tau)
\eeq
where $\tau$ is a two-form on $M_4$. The Maxwell equation can then be solved as
\beq\label{max5}
\Delta_4\tau =  \left(\frac{z}{2}\right)\left(\frac{z}{2}+2\right)\tau \, , \quad \text{where}\quad d\star_4\tau=0
\eeq
where $\Delta_4$ is the Laplace operator of $M_4$ acting here on transverse two-forms. The flux now includes the term
\beq
G \supset  d(r^{z/2}dx^+\wedge\tau) \equiv d(A^{Sch}\wedge\tau)
\eeq
which takes the usual form. The Einstein equation, upon writing $h=r^{z-1}q(M_4)$ in accordance with scale-invariance, is
\beq\label{ein5}
\nabla^2_4q + (z-1)(z+2)q = -\left[\left(\frac{z}{2}\right)^2|\tau|_4^2 + |d\tau|^2_4\right]
\eeq

Note that the case $z=-4$ corresponds to the case of harmonic $\tau$, and so we recover our earlier result as one of many.

The existence of a solution is contingent on solution of the two equations (\ref{max5}) and (\ref{ein5}). For any $M_4$, the Maxwell equation can always be solved to give a spectrum of $z$ unbounded from above. Let us take this to define $z$. So the only way these solutions do not exist is if, for some such $z$, the scalar Laplacian of the Einstein equation admits a homogeneous solution and the quadratic source term includes this mode; this follows our earlier discussion in subsection 2.3.

But it is clear that there is at least some $M_4$ for which there are solutions. Indeed, this is true of the most straightforward choice $M_4 = S^4$, because its scalar and two-form spectra are not aligned so as to allow homogeneous spin-2 solutions for any $z$. On $S^4$, the spectra of functions and transverse two-forms are (e.g. \cite{Elizalde:1996nb})
\beq
\nabla^2_{S^4} q = -l(l+3)q  \, , \quad \Delta_{S^4}\tau = (l+2)(l+3)\tau
\eeq
where $l$ is a non-negative integer. Immediately we see that $z$ will not be rational; indeed, the solution is
\beq
z = -2+4\sqrt{(l+2)(l+3)+1}
\eeq
where we took the positive branch. This gives irrational scalar eigenvalues, as
\beq
(z-1)(z+2) = 4\sqrt{(l+2)(l+3)+1}\left(4\sqrt{(l+2)(l+3)+1}-3\right)
\eeq
So these are solutions on $S^4$, and the solution of the Einstein equation for $q$ will be some linear combination of scalar harmonics.

These solutions are evidently quadratic completions of linearized vector solutions, and are presumably ``hiding'' somewhere in the spectrum of KK fields on AdS$_7 \times M_4$. It is perhaps instructive to plug $z$ into the effective massive vector relation $(mL)^2 = z(z+4)$, which gives integer masses
\beq
\begin{split}
m^2 &= 3+4(l+2)(l+3) \\
\end{split}
\eeq
\subsection{A Lif$_6(z=2)$ solution}

We present this solution as a dimensional reduction of a Sch$_7(z=0)$ solution in the manner first studied by \cite{koush, gauntlif}, analogous to the Lif$_4$ and Lif$_3$ solutions from D3 and M2 branes. All of these, including the present solution, have dynamical exponent $z=2$.

Working once more from the ansatz (\ref{m5ansatz}), we substitute $h=r^{-1}q$, i.e. $z=0$, to give
\beq
\begin{split}
ds^2 & = q(dx^+)^2+r(2dx^+ dx^-  +d\textbf{x}^2) +\frac{dr^2}{r^2} + ds^2(M_4) \\
G& = \star_{11}(3r^2dx^+ \wedge dx^- \wedge d\textbf{x} \wedge dr) +dx^+ \wedge V \\
\end{split}
\eeq
The field equations are
\beq
\begin{split}
&dx^+\wedge d\star_{X_5}\left(Vr^3\right) = 0\\
&dx^+\wedge dV = 0 \\
&r^{-6}(\nabla^2_4q -2q) = -|V|^2_{X_5} \\
\end{split}
\eeq
where we have retained the differentials $dx^+$ because $x^+$ no longer scales, and so the various fields can have $x^+$ dependence consistent with the field equations and scale invariance.

Consider the gauge field equations first. If we restrict $V$ to live on $M_4$ and allow functional dependence on $x^+$ (as in the D3 and M2 brane cases), this will not lead to a solution: $V$ is required by the field equations to be proportional to a harmonic three-form on $M_4$, which does not exist assuming that $M_4$ has at least one continuous isometry.

Instead, we write a more general ansatz for $V$ consistent with symmetry:
\beq
V=A(x^+)\omega + B(x^+) \frac{dr}{r}\wedge\tau
\eeq
where $\omega$ and $\tau$ are a three-form and two-form on $M_4$, respectively. Plugging this into the gauge field equations gives
\beq
\begin{split}
dx^+ \wedge dV&= dx^+ \wedge(A  d\omega - B  \frac{dr}{r} \wedge d\tau) = 0\\
dx^+ \wedge d\star_{X_5}(r^3V) &= dx^+ \wedge (Ardr \wedge d\star_4\omega + 2Brdr \wedge \star_4\tau + Br^2d\star_4\tau) = 0\\
\end{split}
\eeq
Note that we require $A,B \neq 0$. The solution to these equations is
\beq
\begin{split}
\Delta_4\tau&=0\\
d\omega&=0\\
dx^+ \wedge d\star_4\omega &= dx^+ \wedge \left(-2\frac{B(x^+)}{A(x^+)}\star_4\tau\right)\\
\end{split}
\eeq
So $M_4$ must have $b_2\neq0$, reminiscent of the fact that analogous solutions for the M2 brane \cite{gauntlif} require $SE_7$ to support harmonic 3-forms.

The Einstein equation for this solution then reads
\beq
\nabla^2_4q -2q = -\left[A^2|\omega|^2_4+ B^2|\tau|^2_4\right]
\eeq
Expanding $|\omega|^2_4$ and $|\tau|^2_4$ in $M_4$ harmonics, the solution for $q$ is a linear combination of such harmonics with $x^+$-dependent coefficients,
\beq
q = \sum_{n}C_{i_1i_2...i_n}(x^+)Y^{i_1i_2...i_n}(M_4)
\eeq

To summarize, the solution is
\beq
\begin{split}
ds^2 & = q(dx^+)^2+r(2dx^+ dx^-  +d\textbf{x}^2) +\frac{dr^2}{r^2} + ds^2(M_4) \\
G& = \star_{11}(3r^2dx^+ \wedge dx^- \wedge d\textbf{x} \wedge dr) +dx^+ \wedge V \\
\end{split}
\eeq
with fields $q$ and $V$ obeying
\beq
\begin{split}
V &= A(x^+)\omega + B(x^+) \frac{dr}{r}\wedge\tau \, , \quad \text{where}\quad\\
\Delta_4\tau&=0 \, , \quad d\omega=0 \, , \quad dx^+ \wedge d\star_4\omega = dx^+\wedge\left(-\frac{2B}{A}\star_4\tau\right)\\
\end{split}
\eeq
and
\beq
\nabla^2_4q -2q = -\left[A^2(x^+)|\omega|^2_4+ B^2(x^+)|\tau|^2_4\right]
\eeq

Following \cite{koush, gauntlif}, we can write the metric as a circle fibration over a Lif$_6(z=2)$ solution as follows:
\beq
ds^2  = q(dx^++\frac{r}{q}dx^-)^2-\frac{r^2}{q}(dx^-)^2+r(2dx^+ dx^-  +d\textbf{x}^2) +\frac{dr^2}{r^2} + ds^2(M_4)
\eeq
Identifying $x^-$ with the time coordinate and $x^+$ with a compact angle, and restricting to $q \geq 0$, this solution has Lif$_6(z=2)$ symmetry, with the electric U(1) gauge field given by $A = \frac{r}{q}dx^- = \frac{r}{q}dt$. The fact that $z=2$ is clear upon noting that time scales with twice the power of space under dilatations.

\section{Details of Sch$_3$ solutions from D1-D5}

\subsection{AdS$_3 \times S^3 \times M_4$ compactification spectrum}

The compactification of the AdS$_3 \times S^3 \times K3$ solution down to three dimensions was done in \cite{sezgin}. The authors begin from the chiral $d=6, \mathcal{N}=(2,0)$ theory and reduce on the sphere. The matter content of this theory is comprised of the graviton supermultiplet -- which contains five self-dual two-form tensor fields -- coupled to 21 tensor multiplets, each of which contains a single anti-self-dual tensor field. These tensor fields, upon reduction on $S^3$, produce the $d=3$ vectors we will use, and descend from the type IIB complex three-form and five-form fluxes. The nonchiral $\mathcal{N}=(2,2)$ theory which one gets from replacing K3 by $T^4$ has an equal number (five) of self-dual and anti-self-dual tensor fields; but since the reduction of bosonic fields on $S^3$ actually is independent of the number of matter multiplets $n_T$, we need not specify which $d=6$ supergravity we work with.

The reduction on $S^3$ gives rise to five KK towers of vectors: three of these contain mixing between the two-forms and the metric, but two are diagonal. We will refer the reader to \cite{sezgin} for details, and just present the results which we need.

Of the five self-dual two-forms in the $d=6$ theory, one of these descends from the RR two-form $C_2$. In the D1-D5 background, $C_2$ has components turned on that gives a self-dual field strength in six dimensions, as
\beq
F_3=dC_2 = 2(1+\star_6)d\Omega_3
\eeq
so we can isolate one of the five self-dual two-forms as sourcing the AdS$_3 \times S^3$ geometry.\footnote{S-duality rotations in $D=10$ rotate the source terms between the RR and NS-NS sectors.} We borrow the following notation from \cite{sezgin}:
\begin{itemize}
\item $B_{\mu\nu}^5$ is the two-form that sources the D1-D5 background, that is, the self-dual descendant of $C_2$.
\item $B_{\mu\nu}^{\underline{i}}$, where $\underline{i}=1\ldots4$, denotes the remainder of the self-dual two-forms.
\item $B_{\mu\nu}^r$, where $r=1\ldots n_T$, denotes the anti-self-dual two-forms
\end{itemize}

Upon perturbing around this background, there are vector components denoted $b_{\mu a}^5, b_{\mu a}^{\underline{i}}, b_{\mu a}^r$, respectively, where $a=1,2,3$ denotes an $S^3$ coordinate. There are also vectors from components of the metric perturbations with one leg along $S^3$, which are denoted $K_{\mu}$, and of course spin-2 components from the metric with both legs along AdS$_3$.

With this in hand, we delineate the KK tower structure. The three mixed vector towers involve mixing between $K_{\mu}$ and $b_{\mu a}^5$, that is, between the metric and the vector perturbations of the source field. One diagonal vector tower, $b_{\mu a}^{\underline{i}}$, comes from the remaining four self-dual tensor fields; the other, $b_{\mu a}^r$, comes from all of the anti-self-dual tensor fields. There is, of course, also a spin-2 tower of fields.

The only information about these towers which we need to determine the masses is their conformal weights. States of the $D=1+1$ CFT dual to AdS$_3$ are classified by left- and right-moving conformal weights, denoted $(h,\overline{h})$, dual to charges under the $SL(2,R)_L \times SL(2,R)_R$ global symmetry of AdS$_3$. The relation between these weights and the bulk mass of a $p$-form field is
\beq
(mL)^2 = (h+\bar{h}-p)(h+\bar{h}+p-2)
\eeq

The three mixed vector towers have conformal weights
\beq\label{conformalw}
(h,\overline{h}) = \left(\frac{l+x\pm1}{2},  \frac{l+x\mp1}{2}\right)
\eeq
where $x=1,3,5$, and integer $l\geq 0$. This gives rise to masses\footnote{This is actually a bit subtle. The $x=1$ vector branch resides in the spin-2 supermultiplet. The choice $l=0$ vectors are part of a non-propagating supergravity multiplet. They  can be gauged away in the three-dimensional theory, but in formulating our solutions in $D=10$, we will not see this difference manifest. We treat this solution with caution.}, and hence values of $z$,
\beq
(mL)^2=z^2=(l+x-1)^2
\eeq

Both diagonal vector towers have
\beq
(h,\overline{h}) = \left(\frac{l+3\pm1}{2},  \frac{l+3\mp1}{2}\right)
\eeq
again with integer $l \geq 0$, and hence
\beq
(mL)^2=z^2=(l+2)^2
\eeq

Lastly, the spin-2 fields have
\beq
(h,\overline{h}) = \left(\frac{l+2\pm1}{2},  \frac{l+2\mp1}{2}\right)
\eeq
again with integer $l \geq 0$.\footnote{We have included the spin-2 field at $l=0$ which is part of the aforementioned nonpropagating multiplet, so we should beware to discount the resulting Sch$_3$ solution.} The spin-2 mass is defined as
\beq
(mL)^2 = (h+\bar{h})(h+\bar{h}-2)
\eeq
so we expect a spectrum
\beq
(mL)^2 = (2z-2)2z=l(l+2)
\eeq

All of these KK states are organized into representations of the global symmetries of the $d=6$ theory, so $n_T$ determines, in part, the multiplicity of states at a given level. This enters into our Schr\"odinger constructions as determining the degeneracy of solutions. \\

We now have a guide to the construction of the $D=10$ solutions. In particular, we cannot turn on self-dual vector components of $C_2$ without turning on components of the metric with one leg along the sphere. But we can turn on an anti-self-dual vector component of $C_2$ by itself, as well as both self-dual and anti-self-dual vector components of $B_2$ and $A_4$, as these comprise all diagonal vectors $b_{\mu a}^{\underline{i}}, b_{\mu a}^r$. Furthermore, when we take the one-form on $S^3$ to be Killing, we expect to reproduce the TsT-transformed D1-D5 system \cite{oz}, in analogy with the D3-brane construction. We already have evidence for this, because the lowest rung of the diagonal solutions has $z=2$.

Let us now provide complementary material to the solutions in the main text.
\subsection{Diagonal vectors}
\subsubsection{Anti-self-dual RR two-form charge:}
In this case, the Maxwell and Bianchi equations reduce to\footnote{In the remainder of the appendix, we refer to the complex three-form $G_3 \equiv G$.}
\beq
dG = G \wedge G = G \wedge \star_{10} G = d\star_{10} G = 0
\eeq

It is easy to show that
\beq
\star_{10}(iG) = [2(1+\star_{6})d\Omega_3 - (1-\star_6)dx^+ \wedge d(r^z \beta)]\wedge dVol_4
\eeq
where $dVol_4$ is the invariant volume form of $M_4$, so a $\star_{10}$ is essentially equivalent to a $\star_6$. Then $G \wedge G = G \wedge \star_{10} G = 0$ by inspection: since the new component and its Hodge dual both have one leg along $dx^+$ and at least one leg along $S^3$, their wedge product with the background components and with each other will vanish.

The equations $dG = d\star_{10}G = 0$ will give us an eigenvalue equation for the one-form $\beta$. The anti-self-duality of $\delta F_3$ is required for the Einstein equations in the $(+a)$ directions to be trivially satisfied.

This class of solutions includes the TsT-transformed D1-D5 system, in which $q$ is constant, $z=2$, and $\sigma=\eta$, the Reeb Killing vector on $S^3$. From the Einstein equation (\ref{d1d5eqns}), this requires
\beq
\Lambda(\eta)|_{z=2} =4|\eta|^2 + |d\eta|^2 = \text{Constant}
\eeq
This was explicitly shown to be true for the TsT-transformed D3-brane in \cite{bobev}, where $\eta$ lived on $S^5$ instead and was hence $SO(6)$-valued. In fact, this statement is true for any Killing vector of $SO(N)$.

To summarize, the TsT-transformed brane solutions occupy a tidy niche in these constructions: they correspond to turning on the NS-NS two-form with the leg along the compact space given by the Reeb vector of that space. The resulting metrics are direct products.

\subsubsection{NS-NS two-form charge:}

This solution is nearly the same as the previous one.
\subsubsection{RR four-form charge:}
We note that the self-duality ($\star_4\alpha = \alpha$) or anti-self-duality ($\star_4\alpha = -\alpha$) of $\alpha$ with respect to the Euclidean metric on $M_4$ translates into a concordant statement about the $d=6$ tensor fields' self-duality property. We calculate
\beq
\star_{10}((1\pm\star_6)dx^+\wedge d(r^z \beta) \wedge \alpha) = (\star_6 \pm 1)(dx^+ \wedge d(r^z \beta) \wedge \star_4\alpha
\eeq
so that
\beq
F_5 = \frac{(1\pm\star_6)}{\sqrt{2}}dx^+\wedge d(r^z \beta) \wedge \alpha+ \frac{(\star_6 \pm 1)}{\sqrt{2}}dx^+ \wedge d(r^z \beta) \wedge \star_4\alpha
\eeq

If $\star_4\alpha=\alpha$, then we must choose the upper sign -- corresponding to self-duality in $d=6$ -- otherwise $F_5=0$. If $\star_4\alpha=-\alpha$, then we must choose the lower sign -- corresponding to anti-self-duality in $d=6$ -- otherwise $F_5=0$. The expression for $F_5$ becomes
\beq\label{f5}
F_5 = \sqrt{2}(1\pm \star_6)dx^+ \wedge d(r^z\beta)\wedge \alpha
\eeq
and the origin of the $d=6$ tensor fields is clear, as is their multiplicity: the number of cycles which $\alpha$ can wrap gives the degeneracy of such fields.

The Maxwell equations and Bianchi identities are
\beq
\begin{split}
G \wedge \star_{10} G &= dG = 0\\
dF_5 &= \frac{i}{2}G\wedge G^{*} \\
d\star_{10}G &= -\frac{i}{6}G \wedge F_5 \\
\end{split}
\eeq
Because we haven't added any $G$-flux, we have, from the D1-D5 solution,
\beq
G\wedge G^{*} = G \wedge \star_{10}G = d\star_{10}G = dG = 0
\eeq
so we must show that
\beq
G \wedge F_5 = dF_5 = 0
\eeq

The first of these is trivial, based on a previous argument. Imposing $dF_5=0$  gives us the eigenvalue equation.
\subsection{Mixed vectors}
For this subsection, we will `zoom out' from the near-horizon limit of the D1-D5 system and consider the metric of the branes sitting in flat space: that is, we perturb a metric of the form
\beq
ds^2 = \Phi^{-1}(2dx^+dx^-) + \Phi \,ds^2(\mathbb{R}^4) + ds^2(M_4)
\eeq
where the D5 brane wraps $M_4$ with line element $ds^2(M_4)$, and the D1 and D5 intersection along the $x^{\pm}$ directions sits transverse to $\mathbb{R}^4$. This is as we did for the M5 brane. It should be clear that all of the results of the previous subsection for the diagonal vectors could have been found by starting from this metric, and taking $\Phi = r^{-2}$ to be harmonic on $\mathbb{R}^4$ as dictated by the Maxwell equations. We find it instructive to work in this context here, as it will give us a better guide to what goes wrong when we make our initial ansatz.

Informed by the D3 and M5 examples of mixed vector solutions, one might make an ansatz of the form
\beq\label{cansatz}
\begin{split}
ds^2 &= \Phi^{-1}(2dx^+(dx^-+C)) + \Phi\, ds^2(\mathbb{R}^4) + ds^2(M_4) \\
iG &= -(1+\star_6)dx^+ \wedge dx^- \wedge d\Phi^{-1}+(1+\star_6)(dx^+ \wedge d(\Phi^{-1}C)) \\
\end{split}
\eeq

We work in the orthonormal frame
\beq
\begin{split}
e^1 &= dx^+ \\
e^2 &= 2\Phi^{-1}(dx^-+C) \\
e^a &= \sqrt{\Phi} dx^a \, , \, a=3\ldots 6 \\
e^i \cdot e^i &= ds^2(M_4) \\
\end{split}
\eeq
so the metric is given as
\beq
ds^2 = e^1e^2 + e^a \cdot e^a + e^i \cdot e^i
\eeq
The flux ansatz is then
\beq
iG = (1+\star_6)(\frac{1}{2\Phi}e^1 \wedge e^2 \wedge  d\Phi+\frac{1}{\Phi}e^1\wedge  dC)
\eeq

Both the Bianchi and Maxwell equations give
\beq
d\star_{\mathbb{R}^4}d\Phi + d\left(\frac{e^1\wedge \star_{\mathbb{R}^4}dC}{\Phi}\right)=0
\eeq
If we wish to assign the usual $\Phi=r^{-2}$ so as to see the AdS$_3$ region, the second term must vanish. But this is not the equation we want; specifically, the factor of $\Phi$ precludes the usual progression toward an eigenvalue of co-closed one-forms on $S^3$. More to the point, the Einstein equations demand $dC=0$.

If, however, we add a factor of $1/2$ to the flux ansatz in the orthonormal frame as
\beq
iG = (1+\star_6)(\frac{1}{2\Phi}e^1 \wedge e^2 \wedge  d\Phi+\frac{1}{2\Phi}e^1\wedge  dC)
\eeq
then this problem is remedied: the new reduced Maxwell and Bianchi equation is\\
\beq
d\star_{\mathbb{R}^4}d\Phi + \frac{1}{2}\left[d\left(\frac{e^1\wedge \star_{\mathbb{R}^4}dC}{\Phi}\right)+ d\left(\frac{e^1\wedge dC}{\Phi}\right)\right]=0
\eeq\\
If we demand that the last two terms vanish as a pair, we are led to the condition
\beq
\star_{\mathbb{R}^4}dC = -dC \quad \Rightarrow \quad d\star_{\mathbb{R}^4}dC=0
\eeq
That is, $dC$ must be anti-self-dual on $\mathbb{R}^4$. The Einstein equations demand nothing more.

Making the usual substitutions
\beq
\begin{split}
\Phi&=r^{-2}\\
C&=r^{z-2}\beta\\
\end{split}
\eeq
we are led to the eigenvalue equation
\beq\label{s3}
\Delta_{3}\beta = (z-2)^2\beta \, , \quad\text{where}\quad  d\star_{3}\beta=0
\eeq
This gives us Schr\"odinger solutions for two mixed vector branches, with $z=-l, l+4$.

\section{Details of Sch$_4$ solutions from wrapped M5}

The field equations of the $d=4$ theory are rather long; we refer the reader to \cite{wrappedm5}, specifically equations (A.11)$-$(A.24). In the field variables given earlier, one begins by making the following assignments:
\beq
\begin{split}
\beta= \theta_{\alpha}&= \chi_{\alpha}=h_3^{\alpha}=0\\
\T_{\alpha\beta} &= \delta_{\alpha\beta}\\
\lambda,\phi &= \text{constant}\\
B_2 &= \gamma\star dC_1\\
F_2 &\neq 0\\
\end{split}
\eeq
where $\gamma$ is constant. We also impose the null conditions $C_1 \wedge \star  C_1 = B_2 \wedge \star B_2 = F_2 \wedge \star F_2 = 0$, as we are looking for a solution to the field equations in which $F_2$ is also a function of the field strength $B_2 \sim dC_1$.

One finds that $l=-1$ is required for consistency, and that there are exactly two real choices for the values of the pair $(\phi,\lambda)$, namely those given as part of the AdS solutions above. Plugging through the equations, one finds that $F_2$ is proportional to $B_2$ and all other Lagrangian parameters $(L,g,m^2)$ are defined as in the above solutions, subject to the extra conditions
\beq
d\star C_1= dC_1 \wedge dC_1 = \star dC_1 \wedge \star dC_1 = 0
\eeq
These amount to a gauge choice and a vanishing of instanton terms, respectively; the Schr\"odinger gauge field satisfies all constraints. So, for the non-supersymmetric choice $e^{\phi}=6^{-1/4}10^{1/5}, e^{\lambda}=10^{1/10}$ for instance, one ends up with an action
\beq
\mathcal{L} = \int d^4x\sqrt{-g}\left(R+\frac{9\sqrt{6}g^2}{5}\right)-\frac{1}{2}\int dC_1 \wedge \star\, dC_1 - \frac{21\sqrt{6}}{25}g^2\int C_1 \wedge \star\, C_1
\eeq
and the Schr\"odinger solution is implicit.

\end{document}